\documentclass{article}
\usepackage{amssymb,color,graphicx}
\usepackage{amssymb}
\usepackage{wrapfig,framed,cancel}
\usepackage[colorlinks=true, pdfstartview=FitV, linkcolor=blue, citecolor=black, urlcolor=blue]{hyperref}
\definecolor{shadecolor}{rgb}{0.93, 0.93, 0.86}
\newtheorem{theorem}{Theorem}[section]
\newtheorem{prop}{Proposition}[section]

\newtheorem{defn}{Definition}[section]
\newtheorem{lemma}{Lemma}[section] 

\newtheorem{cor}{Corollary}[section]

\newtheorem{remark}{Remark}[section]
 
\def\s{\sigma}
\def\br{\begin{remark}}
\def\er{\end{remark}}

\def\C{{\mathbb C}}
\def\P{{\mathbb P}}
\def\eqref#1{ (\ref{#1})}
\def\&{&\hspace{-20pt}}
\def\R{{\mathbb R}}
\def\d{{\rm d}}

\def\1{\mathbf 1}

\def\wh{\widehat}

\def\wt{\widetilde}
\def \pa{\partial}

\def\bea{\begin{eqnarray}}
\def\eea{\end{eqnarray}}
\def\Id{\mathrm{Id}}
\def\Ai{\mathrm{Ai}}
\def\f{\vec{f}}
\def\g{\vec{g}}
\def\F{\vec{F}}
\def\G{\vec{G}}

\def\res{\mathop{{\rm res}}}
\def\Tr{\mathop{{\rm Tr}}}
\def\Ai{\mathrm{Ai}}
\def\T{\mathrm{T}}

\def\ov{\overline}
\def\QED{ {\bf Q.E.D}\par \vskip 4pt}
\newcommand{\be}{\begin{eqnarray}}
\newcommand{\ee}{\end{eqnarray}}
\newcommand{\bes}{\begin{eqnarray*}}
\newcommand{\ees}{\end{eqnarray*}}
\newcommand{\ds}{\displaystyle}

\definecolor{light-blue}{rgb}{0.8,0.85,1}
\definecolor{blue}{rgb}{0,0,1}
\definecolor{red}{rgb}{1,0,0}

\def\le{\left}
\def\ri{\right}
\def\ba{\begin{eqnarray}}
\def\eeq{\end{eqnarray}}

\renewcommand{\theequation}{\arabic{section}.\arabic{equation}}
\makeatletter
\@addtoreset{equation}{section}
\makeatother

\begin{document}

\baselineskip 16pt plus 1pt minus 1pt
\begin{titlepage}
\begin{flushright}
\end{flushright}
\vspace{0.2cm}
\begin{center}
\begin{Large}
\textbf{The transition between the gap probabilities from the Pearcey to the Airy process; a Riemann--Hilbert approach}
\end{Large}

\bigskip
M. Bertola$^{\dagger\ddagger}$\footnote{Work supported in part by the Natural
  Sciences and Engineering Research Council of Canada (NSERC)}\footnote{bertola@crm.umontreal.ca},  
M. Cafasso$^{\dagger\ddagger}$ \footnote{cafasso@crm.umontreal.ca}
\\
\bigskip
\begin{small}
$^{\dagger}$ {\em Centre de recherches math\'ematiques,
Universit\'e de Montr\'eal\\ C.~P.~6128, succ. centre ville, Montr\'eal,
Qu\'ebec, Canada H3C 3J7} \\
\smallskip
$^{\ddagger}$ {\em  Department of Mathematics and
Statistics, Concordia University\\ 1455 de Maisonneuve W., Montr\'eal, Qu\'ebec,
Canada H3G 1M8} 
\\
\end{small}
\end{center}
\bigskip
\begin{center}{\bf Abstract}\\
\end{center}
We consider the gap probability for the Pearcey and Airy processes; we set up a Riemann--Hilbert approach (different from the standard one) whereby the asymptotic analysis for large gap/large time of the Pearcey process is shown to factorize into two independent Airy processes using the Deift--Zhou steepest descent analysis. Additionally we relate the theory of Fredholm determinants of integrable kernels and the theory of isomonodromic tau function. Using the Riemann--Hilbert problem mentioned above we construct a suitable Lax pair formalism for the Pearcey gap probability and re-derive the two nonlinear PDEs recently found and additionally find a third one not reducible to those.
\end{titlepage}
\tableofcontents

\section{Introduction and description of results}

The Airy kernel 
$$
K_{\Ai}(x,y):=\frac{\Ai(x)\Ai'(y)-\Ai(y)\Ai'(x)}{x-y}$$
 has been known since the nineties (see \cite{MehtaBook},\cite{BowickBrezin} and \cite{MooreMatrix}). It was introduced in relation with the study of random matrices. Its determinant, in fact, has been used to describe the statistical behavior of the Gaussian Unitary  Ensemble near the edge of the spectrum. One of its most striking features has been discovered already in 1991 by Tracy and Widom \cite{TW-Airy}. Consider, for simplicity, the operator related to the Airy kernel restricted to the semi-interval $[s,\infty)$. Its Fredholm determinant, depending on the point $s$, can be expressed through the formula
\be
\det\Big(\Id-(K_\Ai)_{|[s,\infty)}\Big)=\exp\Bigg(-\int_s^\infty(x-s)q(x)^2dx\Bigg)
\label{Buscaglione}
\ee
where $q(x)$ is the Hasting-McLeod solution of the Painlev\'e II equation
$$q''=2q^3+xq$$ uniquely determined by its asymptotics at infinity $q(x)\sim\Ai(x)$ (for an alternative proof of this result, based on the theory of solitonic equations, see also \cite{ASvM}). On the other hand , after its introduction in the nineties, mathematicians and physicists discovered that the Airy kernel and the Tracy-Widom distribution (i.e. its determinant) are not related just to GUE but, rather, they are a sort of universal objects appearing in many different statistical models such as, just to cite few of them, one-dimensional non intersecting Brownian motions, random growth models and random partitions (see for instance \cite{TracyWidomDistro},\cite{JohanssonDOP},\cite{BoroOkouOlsh},\cite{AvM-Airy-Sine},\cite{FerrariSpohn})\\
Few years later it appeared that also the Pearcey kernel has similar properties. It was introduced, around 1998, in relation with matrix models with external source \cite{Brezin5},\cite{BleherKuijlaarsIII}, 1-dimensional Brownian motions \cite{BleherKuijlaarsIII} and plane partitions \cite{OkounkovReshetikhin}. Tracy and Widom, in \cite{TracyWidomPearcey}, deduced some differential equations for its determinant and, with a different approach, some other PDEs have been obtained in \cite{AOvM} and \cite{ACvM1}. In particular, in \cite{ACvM1}, a general method to find some PDEs for general $(1,p)$-kernels is described. The Airy and the Pearcey kernel correspond to the case $p=2,3$   while for $p\geq 3$
 it is believed, as reported in \cite{BergereEynard}, that these $(1,p)$-kernels describe, after a proper rescaling, the statistical behavior of multi-matrix models near the edge of the spectrum. 
These kernels are of the form 
\bea
K^{(p)}(x,y):=\frac 1{(2i\pi)^2}\int_{\gamma_1}d\mu\int_{\gamma_2}d\lambda\frac{e^{\Theta_x(\mu)-\Theta_y(\lambda)}}{\lambda-\mu}
\label{Kp}\\
\Theta_x(\lambda):=\frac{\lambda^{p+1}}{p+1}-\tau_{p-1}\frac{\lambda^{p-1}}{p-1}-\ldots-\tau_2\frac{\lambda^2}{2}+\lambda x
\eea
acting on $I:=[a_1,a_2]\cup\ldots\cup[a_{N-1},a_N]$; here the contours of integration $\gamma_1,\gamma_2$ are curves extending to infinity or  formal linear combinations thereof,  and they are chosen so that  the real part of the exponential tends to minus infinity and hence yields absolutely convergent integrals. It is to be remarked --however-- that no independent proof of the total positivity of these kernels exists in the literature and hence it is not clear whether they describe a determinantal point process. Nevertheless the treatment of their Fredholm determinants can be addressed by completely similar methods and hence we find it worthwhile mentioning them here.

Fredholm determinants of the type (\ref{Buscaglione}) with $K_{\mathrm{Ai}}$ replaced by $K^{(p)}$ (\ref{Kp}) are instances of determinants of operators with {\em integrable kernels} in the sense of Its-Izergin-Korepin-Slavnov
 \cite{ItsIzerginKorepinSlavnov}, as we recall presently (a very concise and clear account is given in \cite{ItsHarnad}).
 
  Given a piecewise smooth oriented curve ${\cal C}$ on the complex plane (possibly extending to infinity) and two $p$-vectors $\vec f(\lambda),\vec g(\lambda)\in L^2({\cal C})$ we define the kernel $K$ as
$$K(\lambda,\mu):=\frac{\vec f^{\mathrm T}(\lambda)\vec g(\mu)}{\lambda-\mu}.$$ Since the kernels we will treat will always satisfy this condition we assume that ${\cal C}$ has no self intersections. 
We say that such kernel is integrable if $\vec f^{\mathrm T}(\lambda)\vec g(\lambda)=0$ (so that it is non-singular on the diagonal). We are interested in the operator $K:L^2({\cal C})\rightarrow L^2({\cal C})$ acting on functions via the formula 
$$(Kf)(\lambda)=\int_{\cal C}K(\lambda,\mu)f(\mu)d\mu$$
and, in particular, we are interested in the Fredholm determinant $\det(\Id-K)$. 
The key observation is that, denoting with $\partial$ the differentiation with respect to any auxiliary parameter on which $K$ may depend, we obtain the formula
\be
	\partial \log\det(\Id-K)=-\mathrm{Tr}((\Id+R)\partial K)
\label{resolvent}\ee 
where $R$ is the resolvent operator, defined as $R=(\Id-K)^{-1}K$. Moreover $R$ is again an integrable operator, i.e. 
$$R(\lambda,\mu)=\frac{\vec F^{\mathrm T}(\lambda)\vec G(\mu)}{\lambda-\mu}$$ and $\vec F,\vec G$ can be found solving the following RH problem:
\bea 
		\Gamma_+(\lambda)&\&=\Gamma_-(\lambda)M(\lambda)\quad\lambda\in{\cal C}\label{RH1}\\
		\Gamma(\lambda)&\& = \1+
\mathcal O(\lambda^{-1})\quad\lambda\longrightarrow\infty\label{RH2}\\
	M(\lambda)&\& =\1
{-}
2\pi i\vec f(\lambda)\vec g^{\mathrm T}(\lambda)
	\label{RH3}\\
	 \vec F(\lambda)&\& =\Gamma(\lambda)\vec f(\lambda)\qquad
		\vec G(\lambda) =(\Gamma^{-1})^{\mathrm T}(\lambda)\vec g(\lambda).
\eea
In several cases of interest the Fredholm determinant for such a kernel coincides with the notion of {\em isomonodromic tau function}
introduced in the eighties by Jimbo Miwa and Ueno \cite{JMU1, JMU2, JMU3} to study monodromy-preserving deformations of rational connections on $\P^1$. In particular in \cite{BorodinDeift} the authors proved that this is the case\footnote{More precisely Borodin and Deift proved that the Fredholm determinants they considered are isomonodromic tau functions for a certain system of Schlesinger equations.} for a large class of kernels including the so--called $_2F_1$-kernel, the Jacobi and the Whittaker kernel. Another good host of examples is provided in \cite{ItsHarnad} and --to a certain extent-- Airy and Pearcey and the other to be considered here have a large overlap with the situation addressed ibidem. 

In the paper we will first connect the Fredholm determinants of these integrable kernels in general terms with the tau function associated to Riemann Hilbert problems introduced in \cite{BertolaIsoTau} following  \cite{Malgrange:IsoDef1}, which reduces to the Jimbo-Miwa-Ueno definition of isomonodromic tau function when the relevant RHP comes from the (generalized) monodromy problem of a rational ODE; this is achieved in Sect. \ref{secBuscaglione} and in particular in Thm. \ref{thdetMalgrange}. The approach followed in the previous literature to connect Fredholm determinants of the Airy kernel on a collection of intervals  $I=[a_1,a_2]\cup\ldots\cup[a_{N-1},a_N]$ was to set-up  a RH problem with jumps on $I$ for instance in \cite{HarnadVir}, to study the integrable differential equations related to the Airy kernel from the point of view of isomonodromic deformations. We take a different approach and define a new $(N+1)\times(N+1)$ RH problem with jumps on the contour ${\cal C}:=\gamma_1\cup\gamma_2$. This RH problem is related to a certain operator acting on $L^2({\cal C})$ and, through an appropriate Fourier transform, we prove that such operator has the same determinant as the operators related to the $(1,p)$-kernel.
This RHP is quite suitable for a Deift--Zhou steepest descent analysis in asymptotic regimes.
 As applications of our approach we give one more (quite straightforward) proof of the Tracy-Widom result for the Airy kernel, we study the asymptotics of the Pearcey kernel using the non-linear steepest descent method and we find a Lax Pair for the Pearcey process.

 The Lax formalism is developed in Sect. \ref{PearceyLax}; using the isomonodromic method \cite{JMU1} we re-derive  the recent nonlinear PDEs satisfied by the Fredholm determinant of the Pearcey kernel  for one interval \cite{ACvM1}. We find also a {\em new and independent} PDE (as to be expected) in Prop. \ref{propPearceyPDE}. While the equation itself is rather complicated, the conceptual significance is that the general solution of the three equations depends on a finite number of parameters, rather than functional ones.

The second main goal of this paper is to relate in a precise form the asymptotic of the Pearcey process and see how it ``becomes'' an Airy process; more precisely, using our setup we consider the asymptotic behavior of the Fredholm determinant of the Pearcey kernel when the endpoints of the intervals is very large as the time parameter grows; we show that the determinant is asymptotically factorized into the product of Fredholm determinants of the Airy process.  While this is to be expected on the ground of physical considerations, we believe this is the first mathematically rigorous proof of this factorization; {a similar but simpler case has been previously studied in \cite{ACvM2} using the ordinary steepest descent method instead of the non--linear one used here}.

The proof is completely detailed for the case of a single large interval in Theorem \ref{main1}, while the case of several intervals is stated in Theorem \ref{main2} and is not significantly different (as it will be apparent) and hence left to the reader.

\section{Fredholm determinants of integrable kernels and tau functions of Riemann--Hilbert problems}
\label{secBuscaglione}
Following \cite{BertolaIsoTau} one may consider a slightly more general notion of tau-function associated to any Riemann--Hilbert problem (RHP) depending on parameters and  which reduces to that of Jimbo-Miwa-Ueno in case such a RHP coincides with the one associated to a rational ODE. We briefly recall the setup of \cite{BertolaIsoTau}; suppose a RHP is posed on a collection of oriented contours $\mathcal C$ 
\bea
\Gamma_+(\lambda;\vec s) &\& = \Gamma_-(\lambda;\vec s) M(\lambda;\vec s)\\
 \Gamma (\lambda;\vec s) &\&= \1 + \mathcal O(\lambda^{-1}) \ , \ \ \ \lambda\to \infty.
\eea 
where $M(\lambda;\vec s):\mathcal C\to SL_r(\C)$ are some suitably smooth functions of $\lambda$, depending smoothly (analytically) on additional deformation parameters, denoted here generically by $\vec s$.
On the space of these deformation parameters, we introduce the following one-form\footnote{In \cite{BertolaIsoTau} the sign is the opposite, which would lead to the tau function to have poles instead of zeroes. We correct this here.} ({here and below we will denote with $'$ the derivative with respect to $\lambda$)}
\bea
	\omega_M(\partial)&\&:= \int_{\cal C}\mathrm{Tr}\Big(\Gamma_-^{-1}(\lambda)\Gamma'_-(\lambda)\Xi_\partial(\lambda )\Big)\frac{\d \lambda}{2\pi i}
	\label{omegamalgrange}\\
	\Xi_\partial(\lambda)&\& :=\partial M(\lambda)M^{-1}(\lambda)
\eea
The definition (\ref{omegamalgrange}) is posed for arbitrary jump matrices; in the case of the RHP (\ref{RH1})--(\ref{RH2})--(\ref{RH3}) the spontaneous question arises as to whether $\omega_M$ in (\ref{omegamalgrange}) and the Fredholm determinant are related. The answer is positive  within a certain explicit correction term, see Thm. \ref{thdetMalgrange}. 

In \cite{BertolaIsoTau} it was shown that $\omega_M$ is also the logarithmic total differential of the isomonodromic  tau function of Jimbo-Miwa-Ueno (in the cases of RHPs that correspond to rational ODEs).
\begin{shaded}
\begin{theorem}\label{thdetMalgrange}
Let $\vec f(\lambda;\vec s), \vec g(\lambda;\vec s): \C \times S \to \C^r$ and consider the RHP with jumps as in (\ref{RH3}).
Given any vector field $\partial$ in the space of the parameters $S$ of the integrable kernel we have the equality
	\be
		\omega_M(\partial)=\partial\ln\det(\Id-K)
		{-}
		H(M)
	\label{detMalgrange}\ee
	where $\omega_M(\partial)$ is as in (\ref{omegamalgrange}) and 
$$H(M):=H_1(M)-H_2(M)=\int_{\cal C} \Big(\partial\vec f'^\T g+\vec f'^\T\partial\vec g \Big)d\lambda -2\pi i\int_{\cal C} \g^\T \f' 
\partial\g^\T\f d\lambda
$$
where $'$ denotes differentiation with respect to $\lambda$.
\end{theorem}
\end{shaded}
The proof is found in Appendix \ref{AppA}. In the case we will treat in the following sections, moreover, we will have $H(M)=0$.
Hence it is possible to define, up to normalization, the isomonodromic tau function $\tau_{JMU}:=\exp(\int\omega_M)$ and this object, thanks to the previous theorem, will coincide with the Fredholm determinant $\det(\Id-K)$.
\section{The Airy kernel}

\begin{wrapfigure}{r}{0.31\textwidth}
\vspace{-15pt}
\resizebox{0.3\textwidth}{!}{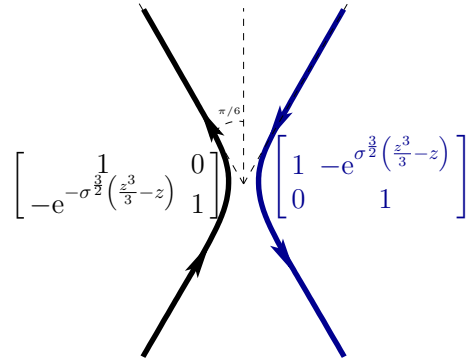}
\caption{The jump matrices for the Hastings--McLeod solution of PII in the rescaled $z$-plane.
}
\label{PII}
\vspace{15pt}
\end{wrapfigure}

We start considering the Airy kernel 
\be
K_\Ai(x,y) := \frac{1}{(2\pi i)^2}\int_{\gamma_R} \d\mu \int_{\gamma_L} \d \lambda \frac {{\rm e}^{\vartheta_x(\mu) - \vartheta_y(\lambda)}}{\lambda - \mu}\ \ \ \vartheta_x(\mu):= \frac {\mu^3}3  - x\mu
\label{Airykernel}\ee
where $\gamma_R$ is a contour in the right  half-plane which extends to infinity along the rays 
$\arg(\lambda) = \pm \frac{i\pi}3 $ and $\gamma_L=-\gamma_R$ (the contours are as in Figure \ref{PII}). We consider the Fredholm determinant $\det(\Id-K_\Ai\chi_I)$ where we denote with the same symbol the Airy kernel and the related operator acting on $L^2(\R)$, $\chi_I$ is the characteristic function of the collection of intervals $I:=[a_1,a_2]\cup[a_3,a_4]\cup\ldots\cup [a_{N-1},a_N]$ for even $N$ and  $I:=[a_1,a_2]\cup[a_3,a_4]\cup\ldots\cup [a_{N},\infty)$ for odd $N$.  The following remark is self evident:
\begin{remark}\label{rem1}
	Let's denote with $K_a(x,y)=K_{\Ai}(x,y)\chi_{[a,\infty)}(x)$. Then we have that 
	\be
		K_{\Ai}(x,y)\chi_I(x)=\sum_{j=1}^N (-1)^{j+1}K_{a_j}(x,y).
	\ee
\end{remark}
Our goal is to setup a RHP associated to the Fredholm determinant of $K(x,y)\chi_I(y)$; of course the interest is somewhat limited since this issue has been thoroughly investigated in the literature (see for instance \cite{DIZ} for the similar case of the sine kernel and more generally  \cite{BorodinDeift}). We point out --however-- that the RHP that we are going to set up is not of the same nature as the natural one considered in \cite{HarnadVir} or in \cite{BorodinDeift}. We will use this formulation later on to investigate the  asymptotic behavior of the Pearcey Fredholm determinant.
\begin{defn}
\label{multiAi}
	Given $I=[a_1,a_2]\cup\ldots\cup[a_{N-1},a_N]$  for even $N$ or $I=[a_1,a_2]\cup\ldots\cup[a_{N},\infty)$ for odd $N$ we define the related AiO-RH problem (standing for Airy--operator Riemann-Hilbert problem)
	\be
	\begin{array}{ccc}
	 	\Gamma_+(\lambda)=\Gamma_-(\lambda)\big(\1
		{-}
		G(\lambda)\big)\qquad \lambda\in\gamma:=\gamma_R\cup\gamma_L\\
		\\
		\Gamma(\lambda)= \1+\mathcal O(\lambda^{-1})\quad \lambda\longrightarrow\infty\\
		\\
		G(\lambda):= 
				 \le[\begin{array}{ccccc}
					0 & {\rm e}^{\vartheta_{a_1}(\lambda)} & - {\rm e}^{\vartheta_{a_2}(\lambda)} & \dots &(-)^{N+1} {\rm e}^{\vartheta_{a_N}(\lambda)}\\
					&0&0&\dots &\\ 
					&&\ddots&&\\
					&&&&\\
					&&&&0
					\end{array}\ri]\chi_{\gamma_R}(\lambda)+
				\le[ \begin{array}{ccccc}
					0&&&&\\
					{\rm e}^{-\vartheta_{a_1}(\lambda)} &0&&&\\
					{\rm e}^{-\vartheta_{a_2}(\lambda)} &0&0&&\\
					&&&\ddots&\\
					{\rm e}^{-\vartheta_{a_N}(\lambda)}&&&&0
					\end{array}\ri]\chi_{\gamma_L}(\lambda)
	\end{array}
	\label{AIRH}
\ee 
	which consists of finding an analytic matrix-valued function $\Gamma(\lambda)$ on $\C\setminus \gamma$ normalized to the identity at infinity and such that the limiting value of $\Gamma(\lambda)$ approaching the contour from the left and from the right ($\Gamma_+(\lambda)$ and $\Gamma_-(\lambda)$ respectively)
	are related through the jump matrix $G(\lambda)$. 
\end{defn}
\br[Solvability of the problem in Def. \ref{multiAi}]
The problem in Def. \ref{multiAi} will be associated to an integrable kernel  in Prop. \ref{31}. The solvability is then equivalent (by the theory of Its-Izergin-Korepin-Slavnov \cite{ItsIzerginKorepinSlavnov}) to the non-vanishing of the corresponding Fredholm determinant. It will be shown in Thm. \ref{KKtilde} that this determinant is precisely the gap-probability of the Airy process, hence (strictly) positive. This guarantees automatically the solvability for real values of the parameters $a_j$. In the case $N=1$ the solvability of the problem in Def. \ref{multiAi} is equivalent to the absence of poles in the Hastings--McLeod solution, shown in \cite{Hastings-McLeod}.
The same remark, {\em mutatis mutandis}, shall apply to the Riemann--Hilbert problem in Def. \ref{defRHPmulti} using Theorem \ref{KPtilde}. 
\er

The RHP of Def. \ref{multiAi} is associated to a certain Fredholm determinant of an integrable operator on $\gamma_R\cup \gamma_L$ as explained in the introduction.
\begin{prop}
\label{31}
	The Riemann-Hilbert problem (\ref{AIRH}) is the RH problem associated to the integrable operator 
	\bea
		{\widetilde K(\lambda,\mu)}&:=& \frac {\vec f^\T(\lambda) \vec g(\mu)}{\lambda-\mu}\\
\vec f(\lambda)&:=&\frac{1}{2\pi i}\le( \le[\begin{array}{c}  
{\rm e}^{\frac {\lambda^3}6}\\
		0\\ \vdots\\ 0
		\end{array} \ri] \chi_{\gamma_R}(\lambda)+ \le[\begin{array}{c}
		0\\ {\rm e}^{-\frac{\lambda^3}6 + a_1\lambda} \\ \vdots\\ {\rm e}^{-\frac{\lambda^3}6 + a_N \lambda}\end{array}\ri] \chi_{\gamma_L}(\lambda)\ri)
\label{Aif}
\\
\vec g(\lambda)&:=&
  \le[\begin{array}{c}
		0\\  {\rm e}^{ \frac {\lambda^3}6 - a_1\lambda} \\ \vdots\\ (-1)^{N+1} {\rm e}^{
		 \frac {\lambda^3}6 - a_{_N}\lambda}  \end{array}\ri] \chi_{\gamma_R}(\lambda)
+
\le[\begin{array}{c}  {\rm e}^{-\frac {\lambda^3}6 }\\
		0\\ \vdots\\ 0
		\end{array} \ri] \chi_{\gamma_L}(\lambda)
\label{Aig}
\eea
\end{prop}
\noindent\emph{Proof:} It is just enough to verify that $\1
{-}
2\pi i \vec f(\lambda)\vec g(\lambda)^{\mathrm T}=\1
{-}
G(\lambda)$.
\QED
\br
\label{important}
The precise form of the contours $\gamma_L$, $\gamma_R$ is not essential; in fact we could replace $\gamma_L$ by the imaginary axis. One can directly show that the Riemann--Hilbert problems obtained by this ``contour deformation'' of $\gamma_L$ to $i\R$ are equivalent. 
\er

\begin{theorem} 
\label{KKtilde}
The following identity holds
	$$\det(\Id-\widetilde K)=\det(\Id-K_{\Ai}\chi_I).$$
where the operator defined by $\widetilde K$ is a trace-class operator on $L^2(\gamma_L\cup \gamma_R)$.
\end{theorem}

\noindent{\bf Proof:} 
We start with observing that any operator on $L^2(\gamma_L \cup \gamma_R) \simeq L^2(\gamma_L) \oplus L^2(\gamma_R) = \mathcal H_1 \oplus \mathcal H_2$ can be written as a  $2\times 2$ matrix  of operators with $(i,j)$ entry given by an operator $\mathcal H_i\to \mathcal H_j$.
Writing out $\wt K$ in full we have 
\be
\wt K(\lambda,\mu) = \frac 1{2i\pi} \frac { {\rm e}^{\frac{\lambda^3-\mu^3}6} \chi_{_{\gamma_L}}(\mu)\chi_{_{\gamma_R}}(\lambda)  
- \sum_{j=1}^{N}  (-)^j {\rm e}^{ \frac {\mu^3 - \lambda^3}6 - a_j \mu
 +a_j\lambda} \chi_{_{\gamma_R}}(\mu)\chi_{_{\gamma_L}}(\lambda)}{\lambda - \mu}
\ee
Define  operators $\mathcal F,\mathcal G_{a}$ as follows
\bea
&& \begin{array}{ccc}
 \mathcal G_a: L^2(\gamma_R)&\longrightarrow & L^2(\gamma_L) \\
f(\mu) & \mapsto & \ds {\rm e}^{a\xi-\frac {\xi^3}6  }\frac{1}{2\pi i}\int_{\gamma_R}{\rm e}^{  \frac{\mu^3}{6} - a\mu} \frac{f(\mu)}{\xi - \mu}\d\mu
\end{array}\\
&&\begin{array}{ccc}
 \mathcal F: L^2(\gamma_L)&\longrightarrow & L^2(\gamma_R)\\
g(\lambda) & \mapsto & \ds \frac{{\rm e}^{\frac {\mu^3}6}}{2\pi i}\int_{\gamma_L}{\rm e}^{-\frac{\lambda^3}6 } \frac{g(\lambda)}{ \mu - \lambda }\d\lambda.
\end{array}\ .
\eea
We will think of these operator as acting on $\mathcal H:=\mathcal H_1\oplus \mathcal H_2$ by extending them trivially to the orthogonal complements of the respective domains. 
The operators $\mathcal G_a$ and $\mathcal F$ are of Hilbert--Schmidt class (HS for short) in $\mathcal H$: this follows from the convergence of the following expressions: 
\be
\|\mathcal G_a\|_{\mathcal H,2}^2 = \int_{\gamma_L}|\d \xi|  \int_{\gamma_R} \!\!\! |\d\mu| \,\frac { {\rm e}^{
2 \Re ( \frac{\mu^3}6 - a\mu - \frac {\xi^3} 6) }}{4\pi^2 |\xi-\mu|^2}<+\infty
\ee
\be
\|\mathcal F \|_{\mathcal H,2}^2 = \int_{\gamma_L}|\d \xi| \int_{\gamma_R} \!\!\! |\d\mu|\,\frac { {\rm e}^{
2 \Re ( \frac{\mu^3}6 - \frac {\xi^3} 6) }}{4\pi^2|\xi-\mu|^2} <+\infty
\ee
\br
\label{def1}
Note  that they would still be of HS class even if we replaced $\gamma_L$ by $i\R$ (which we will do later).
\er
More is true: both $\mathcal G_a$ and $\mathcal F$ are of {\bf trace-class}. To see this it is sufficient to write them as the composition of two HS operators. To achieve this goal we introduce an additional contour $\gamma_0:= i\R + 
\epsilon$ not intersecting either of $\gamma_{L, R}$ and extend the Hilbert space by adding $\wh {\mathcal H}:= L^2(\gamma_0)\oplus \mathcal H$.
Consider now  the two operators on $\wh {\mathcal H}$ defined by 
\be
\begin{array}{c|c}
\mathcal C^{(1)}_a: L^2(\gamma_R)\to  L^2 (\gamma_0) & \mathcal C^{(2)}_a:L^2(\gamma_0)\to L^2(\gamma_L)\\[10pt]
\ds \mathcal C^{(1)}_a (f)(\zeta):= \int_{\gamma_R} \frac{f(\mu) {\rm e}^{\frac{\mu^3}6 - a\mu} \d \mu}{2i\pi(\zeta-\mu)} & \ds \mathcal C^{(2)}_a(h)(\xi):= {\rm e}^{-\frac {\xi^3}6 + a \xi}  \int_{\gamma_0}\frac{ h(\zeta) \d \zeta}{2i\pi(\zeta-\xi)}
\end{array}
\ee
and extended trivially on the orthogonal complements of the respective domains within $\wh {\mathcal H}$.
 They both are HS  in $\wh {\mathcal H}$ because
\be
\le\|\mathcal C^{(1)}_a\ri\|_{\wh {\mathcal H}, 2}^2 =  \int_{\gamma_0}|\d \zeta | \int_{\gamma_R} \!\!\! |\d\mu|\,\frac { {\rm e}^{
2 \Re ( \frac{\mu^3}6 -a \mu) }}{4\pi^2|\zeta-\mu|^2} <+\infty
\qquad  
\le\|\mathcal C^{(2)}_a\ri\|_{\wh {\mathcal H}, 2}^2 =  \int_{\gamma_L}|\d \xi | \int_{\gamma_0} \!\!\! |\d\zeta|\,\frac {
 {\rm e}^{ 2\Re (-\frac {\xi^3}6 + a \xi) }}{4\pi^2|\zeta-\xi|^2} <+\infty\nonumber
\ee
A simple Cauchy residue-computation closing the integration by a big circle to the left  shows
\be
\mathcal G_a f(\xi)= \mathcal C^{(2)}_a\circ \mathcal C^{(1)}_a f(\xi)
\ee
 and hence $\mathcal G_a$ is  the composition of two HS operators and thus   of trace-class.
 For $\mathcal F$ one has to use instead the operators 
\be
\begin{array}{c|c}
\mathcal D^{(1)}: L^2(\gamma_L)\to L^2 (\gamma_0) & \mathcal D^{(2)}:L^2(\gamma_0)\to L^2(\gamma_R)\\[10pt]
\ds \mathcal D^{(1)}(f)(\zeta):=\int_{\gamma_L}  {\rm e}^{-\frac{\lambda^3}6 }\frac{f(\lambda)  \d \lambda}{2i\pi(\lambda-\zeta)} & 
\ds \mathcal D^{(2)}(h)(\mu):= {\rm e}^{\frac {\mu^3}6}  \int_{\gamma_0}\frac{ h(\zeta) \d \zeta}{2i\pi(\zeta-\mu)}
\end{array}
\ee
(both HS)
that realize $\mathcal F = \mathcal D^{(2)} \circ \mathcal D^{(1)}$.
\br
\label{nodef1}
Note that if we replace $\gamma_L$ by $i\R$ it is {\em still true} that $\mathcal G_a = \mathcal C^{(2)}_a \circ \mathcal C^{(1)}_a$, $ \mathcal F =  \mathcal D^{(2)} \circ \mathcal D^{(1)}$ {\bf but} now $\mathcal D^{(1)}$ and $\mathcal C^{(2)}_a$ fail to be HS, and hence we do not know whether $\mathcal G_a, \mathcal F$ are of trace-class between $L^2(\gamma_R)$ and $L^2(i\R)$ (they are  still HS, see Rem. \ref{def1}).
\er
According to the split $\mathcal H= L^2(\gamma_L)\oplus L^2(\gamma_R)$ and using matrix notation, we can write $\det(\Id-\widetilde K)$ as
\be
\det\le[\Id - \le[\begin{array}{c|c}
0 & \sum_{j=1}^{N} (-)^{j+1} \mathcal G_{a_j}\\
 \hline
\mathcal F & 0
\end{array}\ri]\ri] = \det[\Id - \sum_{j=1}^{N} (-1)^{j+1} \mathcal G_{a_j} \circ \mathcal F]
\label{eqA}
\ee
Both determinants are well--defined because of the form $\Id+ $trace class.
This identity between Fredholm determinants of operators follows by multiplying on the left the operator in the left hand side by the operator (with unit determinant)
\be
\Id + \le[\begin{array}{c|c}
0 & \sum_{j=1}^{N} (-)^{j+1}\mathcal G_{a_j}\\
\hline
0 & 0
\end{array}\ri]
\ee
Note that the operator $\mathcal K:= \sum_{j=1}^{N} (-1)^{j+1} \mathcal G_{a_j} \circ \mathcal F$ appearing in the second term of (\ref{eqA}) is an operator acting on $L^2(\gamma_L)$ into itself and with kernel
\be
\mathcal K(\xi, \lambda):=\sum_{j=1}^N (-1)^{j+1} \frac {{\rm e}^{- \frac {\lambda^3+\xi^3}6 + a_j \xi} }{(2i\pi)^2}\int_{\gamma_R} \d\mu \frac{ {\rm e}^{\frac {\mu^3}3 - a_j\mu} }{(\xi-\mu)(\mu-\lambda)}\ ,\ \ \lambda, \xi\in \gamma_L\ .
\label{K0}
\ee
The integral operator on $L^2(\gamma_L)$ defined by $\mathcal K$ can be made to act on $L^2(i\R)$, with the same kernel but where now $\xi,\lambda\in i\R$: let us denote this new operator $\mathcal K_0$. Following Remark \ref{def1} we know that $\mathcal K_0:L^2(i\R)\to L^2(i\R)$ is of trace-class because it is the composition of HS operators. It should be also clear that the Fredholm determinants of $\Id_{\gamma_L} - \mathcal K$ and $\Id_{i\R} - \mathcal K_0$ are the same. Indeed, in the series that computes the two determinants, the contour $\gamma_L$ can be continuously deformed to $i\R$ given the analyticity of the kernel. 
We now conjugate the integral operator on $L^2(i\R)$ defined by $\mathcal K_0$ by the following   Fourier--like unitary operator 
\be
 \begin{array}{rcl}
	\mathcal T: L^2(i\R )&\longrightarrow & L^2(\R) \\
f(\xi) & \mapsto & \ds \frac 1{\sqrt{2i\pi}} \int_{i\R}f(\xi){\rm e}^{\frac{\xi^3}6 - \xi x} \d\xi
\end{array}
\qquad
 \begin{array}{rcl}
	\mathcal T^{-1}: L^2(\R )&\longrightarrow & L^2(i\R) \\
h(x) & \mapsto & \ds \frac { {\rm e}^{-\frac{\xi^3}6} } {\sqrt{2i\pi}} \int_{\R}h(x){\rm e}^{ \xi x} \d x.
\end{array}
\ee
Consider now (the kernel of the operator defined by)  $\mathcal T\circ\mathcal K_0 \circ\mathcal T^{-1}$: each of the terms in its defining sum in eq. (\ref{K0}) gives
 \be
 \mathcal T\circ\mathcal G_a\circ \mathcal F \circ\mathcal T^{-1}=: \mathcal L_{a}(x,y) = \frac{1}{(2\pi i)^2} \int _{i \R}\frac{\d\xi}{2\pi i} {\rm e}^{ \xi  (a-x) }\int_{\gamma_R} \!\!\!\! \d \mu \int_{i\R}\!\!\! \d\lambda\,\,\frac{{\rm e}^{\frac {\mu^3 - \lambda^3} 3 - a\mu }}{(\xi - \mu)(\mu-\lambda)}
{\rm e}^{ y \lambda}
 \ee
Now, if $x>a$ we can close the $\xi$--integration with a big semi-circle in the right half plane, picking up only {\em minus}  the residue at $\mu\in\gamma_R$; viceversa, if $x<a$ we close the $\xi$--integration with a big semi-circle in the {\em left} half plane, which yields zero since there are no singularities within this contour of integration. In summary 
\bea
 \mathcal L_a(x,y) = \left \{
 \begin{array}{cc}
 \ds  
-\frac{1}{(2\pi i)^2}\int_{\gamma_R} \!\!\!\! \d \mu \int_{i\R}\!\!\! \d\lambda\,\,\frac{{\rm e}^{\frac {\mu^3 -  \lambda^3} 3 - x \mu + y \lambda }}{(\mu-\lambda)}
 & x>a\\[20pt]
 \ds
0  & x<a
  \end{array}\qquad =  K_a(x,y)
 \right.
 \label{212}
\eea
This proves that (using Remark \ref{rem1}) that
\be
\mathcal T \circ \mathcal K_0 \circ \mathcal T^{-1} = K_{\Ai} \big|_{I}
\ee
is the Airy kernel restricted to the union of interval $I$, concluding the proof.
\QED

Looking at 
Theorem 
\ref{thdetMalgrange} we verify that in this case  $H(M)\equiv 0$ where $M(\lambda) = \1 + G(\lambda)$ as in Def. \ref{multiAi}; indeed the vectors $\vec f, \vec g$ (\ref{Aif}, \ref{Aig}) satisfy the stronger   identity  $\vec f^T(\lambda;\vec a) \cdot \vec g(\mu; \vec a')\equiv 0$ when $\lambda,\mu$ {\bf both} belong to the same $\gamma_R$ (or $\gamma_L$), respectively; indeed this implies that the additional integrands defining $H(M)$ of formula (\ref{detMalgrange}) are identically zero. Thus we deduce immediately the following
\begin{theorem}\label{Fredholmtau}
The Fredholm determinant $\det(\Id-K\chi_I)$, with $I=[a_1,a_2]\cup\ldots\cup[a_{N-1},a_N]$ for even $N$ or $I=[a_1,a_2]\cup\ldots\cup[a_{N},\infty)$ for odd $N$  is equal to the isomonodromic tau function of the RH problem (\ref{AIRH}) namely
\be
\pa_{a_\ell} \ln \det(\Id-K_{\Ai}\chi_I) = \omega_M(\partial_{a_\ell})&\&:= \int_{\gamma_R\cup \gamma_L}\mathrm{Tr}\Big(\Gamma_-^{-1}(\lambda )\Gamma'_-(\lambda)\Xi_{\partial a_\ell}(\lambda )\Big)\frac{\d\lambda}{2\pi i}\ .
\ee
\end{theorem}
Theorem \ref{Fredholmtau} implies also some more explicit differential identities by using the Miwa-Jimbo-Ueno residue formula; note first that the jump matrices $M$ can be written as 
\bea
M(\lambda;\vec a) = {\rm e}^{T(\lambda)} M_0 {\rm e}^{-T(\lambda)}
\eea
where $M_0$ is a {\em constant} matrix (consisting of only $\pm 1$ and $0$) and 

\bea
T(\lambda;\vec a) &\& = {\rm diag}\le(T_0,T_1,\dots, T_{
N}\ri)\\
T_0 &\& := \frac 1{
{N}+1} \sum_{j=1}^{
N} \vartheta_{a_j}\qquad 
T_\ell := T_0 - \vartheta_{a_\ell} 
\eea
The matrix $\ds \Psi(\lambda;\vec a):= \Gamma(\lambda;\vec a)\, {\rm e}^{T(\lambda;\vec a)} $
 solves a RHP with constant jumps and hence is (sectionally) a solution to a polynomial ODE.  It was shown in \cite{BertolaIsoTau} that (adapting to the situation at hand)
\be
 \int_{\gamma_R\cup \gamma_L}\mathrm{Tr}\Big(\Gamma_-^{-1}(\lambda)\Gamma'_-(\lambda)\Xi_{\partial_{a_\ell}}(\lambda)\Big)\frac{\d\lambda}{2\pi i} = -  \res_{\lambda=\infty} {\rm Tr} \le(\Gamma^{-1}(\lambda) \Gamma'(\lambda) \pa_{a_\ell} T \ri)\label{219}
\ee
We then find 
\begin{prop}
\label{Freddef}
The Fredholm determinant $\det(\Id-K\chi_I)$, with $I=[a_1,a_2]\cup\ldots\cup[a_{N-1},a_N]$   for even $N$ or $I=[a_1,a_2]\cup\ldots\cup[a_{N},\infty)$ for odd $N$  satisfies
\be
\pa_{a_\ell} \ln \det(\Id-K\chi_I) =-\Gamma_{1;\ell+1,\ell+1},
\ee
where $\Gamma_1:= \lim_{\lambda \to \infty} \lambda \le(\Gamma(\lambda) -\1\ri)$ and $\Gamma(\lambda)$ is the solution of the RHP in Def. \ref{multiAi}.
\end{prop}
{\bf Proof.} This follows directly from formula (\ref{219})  and from 
\be
\frac{\pa}{\pa a_\ell} T(\lambda;\vec a )  = -\lambda\le(\frac 1 {
N+1
} \1 - E_{\ell+1,\ell+1}\ri)\label{221}\ .
\ee
Indeed, plugging (\ref{221}) into (\ref{219}) we find 
\be
\pa_{a_\ell} \ln \det(\Id-K\chi_I) = \frac 1{N+1} \Tr \Gamma_1 - \Gamma_{1;\ell+1,\ell+1}.
\ee
Since $\det \Gamma(\lambda)\equiv 1$ it easily follows that $\Tr \Gamma_1 = 0$, whence the proof. \QED

Now let's consider the case $N=1$, i.e. the case in which we study the Airy kernel on the semi-infinite interval $[s,\infty)$. In this case it is immediate to observe that the RH problem \ref{AIRH} is nothing but the Riemann-Hilbert problem for the so-called Hasting Mc-Leod solution of the PII equation
\be
\begin{array}{ccc}
	q''(s)&=&2q^3(s)+sq(s)\\
	\\
	q(s)&\sim&\Ai(s),\quad s\longrightarrow +\infty
\end{array}
\label{HM}\ee
(see for instance \cite{ItsKapaevFokasBook}; actually in the standard form the phase should be $\vartheta_x(\mu)= i\frac {4\mu^3}3  + i x\mu$ but the two different formulations are equivalent).
Using our approach we can give a proof of the connection between the Airy kernel and Painlev\'e II equation, as stated for the first time by Tracy and Widom \cite{TW-Airy}; note that this proof has purely academic relevance inasmuch as it is of different nature from the original one.

\begin{cor}\label{TWdistribution}
	Consider the semi-interval $[s,\infty)=I$, $N=1$; then we have that 
	\be
		\det(\Id-K_{[s,\infty)})=\exp\Bigg(-\int_s^\infty(x-s)q(x)^2dx\Bigg)
	\label{TW}\ee
	where $q(x)$ is the Hasting Mc-Leod solution (\ref{HM}) of the Painlev\'e II equation.
\end{cor}
\noindent{\bf Proof.} 
The RHP \ref{AIRH} for $N=1$ implies the symmetry
\be
\Gamma(\lambda) = \sigma_1 \Gamma(-\lambda) \sigma_1\ ,
\ee
since the jump matrices have the same symmetry. In particular
\be
F_1(s)&:=&\lim_{\lambda\to \infty} \lambda\le(\Gamma(s,\lambda)-\1\ri)=
{p(s)\s_3 + iq(s) \s_2}
\label{F1}
\ee
On the other hand we have already proven that the determinant is the isomonodromic tau function and from Proposition \ref{Freddef} we find 
\be
\partial_s\log\det(\Id-K_{[s,\infty)})=p(s) .
\ee
{
Now it is well known and easy to prove (see again \cite{ItsKapaevFokasBook})  that the matrix $\Psi(\lambda):=\Gamma(\lambda)\exp(\frac 1 2 \vartheta_x \sigma_3)$) satisfies a RH problem with constant jump and, consequently, the Lax system
\be
	\pa_s\Psi(\lambda;s)=U(\lambda;s)\Psi(\lambda,x)\quad\quad\quad \pa_\lambda\Psi(\lambda;s)=V(\lambda;s)\Psi(\lambda;s)
\label{LaxPII}\ee
with
\be
U(\lambda;s)&\& :=
{-\frac \lambda 2 \s_3 - q(s)\s_1}\ ,
\quad 
V(\lambda;s) :=
{-\lambda U(\lambda;s) + \le(p'(s)-\frac \lambda 2 \ri)\s_3 +i q'(s)\s_2}
\ee
Using the compatibility condition between the two equations above one can show that $p'(x) =-q^2(x)$ and $q(x)$ solves the PII equation}. Integrating twice we get the formula (\ref{TW}) while the asymptotic of $q(x)$ can be deduced from its Stokes parameters as done in \cite{ItsIzerginKorepinSlavnov} and recalled just here below.\QED

\subsection{Asymptotic behavior of the Hastings--McLeod auxiliary matrix}\label{auxiliarymatrix}
In the following we need some information on the behavior for $\sigma \to +\infty$ of the AiO-RH problem (\ref{AIRH}) with phase $\vartheta_\sigma(\zeta)$. The AiO-RH problem for N=1 is nothing but the well known RH problem for the Hasting-Mc Leod solution of PII as discussed in Corollary \ref{TWdistribution}.  We need some simple estimates for the matrix $\Gamma_{Ai}(\zeta;\sigma)$ as $\sigma\to +\infty$ and $|\zeta|\to\infty$. To this end we introduce the scaling $\zeta =\sqrt {\sigma} z$. The saddle point for the phase functions are:
\be
\vartheta_\sigma(\sqrt{\s} z) = \s^\frac 32\le(\frac {z^3}3 - z\ri) \ \ \ z_c = \pm 1 \ \ \ \vartheta_c = \mp \frac 23 \sigma^{\frac 32}.
\label{337}
\ee

The contours of the jumps can be continuously deformed so as to pass through the saddle points $z_{1,2} := \pm 1$ of the phase on the regular steepest descent contours $\Im \le( {z^3}/3-z\ri) = 0$.
It appears from (\ref{337}) and the shape of the jumps (Fig. \ref{PII}) that both jump matrices are uniformly close to the identity in any $L^p$ norm (including $p=\infty$) with all these norms being  $\mathcal O({\rm e}^{-\frac 23 \s^{\frac 32}})$ by simple steepest-descent estimates. Since the jumps are analytic and the norm estimates hold also for deformed contours going to infinity within suitable open sectors ($|\arg(z)| \in \le(\frac \pi 6 + \epsilon, \frac \pi 2-\epsilon \ri)$ for $\gamma_R$, for example) the small norm theorem guarantees  (for large $\s>0$)
\be
\le\|\Gamma_{\Ai}\le(\zeta;\sigma\ri) - \1 \ri\| \leq \frac {C\, {\rm e}^{-\frac 2 3 \sigma^{\frac 32}} }{1+\frac{|\zeta|}{\sqrt{\s}}}\leq \frac {C\sqrt{\s} {\rm e}^{-\frac 2 3 \sigma^{\frac 32}} }{1+|\zeta|}\label{smallPII}
\ ,
\ee 
with $C$ some (inessential) constant, independent of $\sigma, \zeta$. What will matter for us is that the $\sigma$-dependence of $\Gamma_{Ai}$ is exponentially small as $\sigma\to +\infty$.
\br
\label{rem36}
It is important (but also simple) that the contours for the jumps of the Hastings--McLeod $\Psi$--functions can be deformed to the rays of angles $\pm \pi/4$.
\er
\section{The Pearcey kernel}
\label{PearceyKernel}
In this section we perform the same analysis as the one for the Airy kernel for the case of the Pearcey kernel
\bea
K_{_P}(x,y,\tau) &:=&\frac{1}{(2\pi i)^2} \int_{\gamma_L\cup\gamma_R}\hspace{-10pt} \d\mu \int_{i\R} \d \lambda \frac {{\rm e}^{\Theta_x(\mu) - \Theta_y(\lambda)}}{\lambda - \mu} \label{PKern}\\
\Theta_x(\mu)&:=& \frac {\mu^4}4 - \frac {\tau}2 \mu^2  -  x\mu.\label{ThetaP}
\eea

\begin{wrapfigure}{r}{0.3\textwidth}
\resizebox{0.3\textwidth}{!}{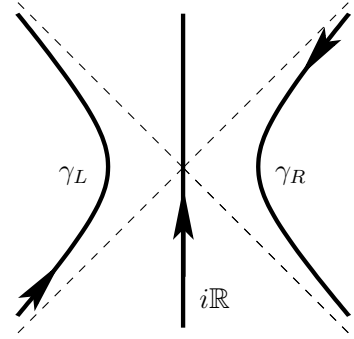}
\caption{The contours for the Pearcey kernel and Riemann--Hilbert problem}
\label{PearceyContours}
\end{wrapfigure}

The contours $\gamma_L,\gamma_R$ are indicated in Fig. \ref{PearceyContours}.
Our aim is to consider the Fredholm determinant $\det(\Id-K_{_P}\chi_I)$ where we denote with the same symbol the Pearcey kernel and the related operator acting on $L^2(\R)$. $\chi_I$ is the characteristic function of the collection of intervals $I:=[a_1,a_2]\cup[a_3,a_4]\cup\ldots\cup [a_{2N-1},a_{2N}]$. 
The relevant RH problem in this case is the following:
\begin{defn}
\label{defRHPmulti}
Given $I=[a_1,a_2]\cup\ldots\cup[a_{2N-1},a_{2N}]$ we define the related PO-RH problem (Pearcey operator Riemann-Hilbert problem)
\bea
\Gamma_+(\lambda)&\& =\Gamma_-(\lambda)\le (\1
{-}
G(\lambda)\ri ),\quad \lambda\in 
\gamma_L\cup \gamma_R\cup i\R  , \\
 \Gamma(\lambda)&\& =\1 + \mathcal O(\lambda^{-1}),\quad \lambda\longrightarrow\infty\\
G(\lambda)&\&:= \!\!\!
				 \le[\begin{array}{cccc}
					0 &\!\!\! {\rm e}^{\Theta_{a_1}(\lambda)} & \!\!\!-{\rm e}^{ \Theta_{a_2}(\lambda)} &\!\!\!\! 
					 \dots(-)^{2N+1}{\rm e}^{\Theta_{a_{2N}}(\lambda)}\\
					&0&\dots&0 \\ 
					&&\ddots&\\
					&&&\\
					&&&0
					\end{array}\ri]
					\!\! \chi_{\gamma_{_R}\cup \gamma_{_L}}(\lambda)\!+\!\!
				\le[ \begin{array}{cccc}
					0&&&\\
					{\rm e}^{-\Theta_{a_1}(\lambda)} &\hspace{-14pt}0&&\\
					\vdots&&\hspace{-10pt}\ddots&\\
					{\rm e}^{-\Theta_{a_{2N}}(\lambda)}&&&0
					\end{array}\ri]\!\!\!\chi_{i\R}(\lambda)
\nonumber\\
					 \label{PRH}
\eea 
	which consists of finding an analytic matrix-valued function $\Gamma(\lambda)$ on $\C\setminus (\gamma_L\cup\gamma_R\cup i\R)$ normalized to the identity at infinity and such that the limiting value of $\Gamma(\lambda)$ approaching the contour from the left and from the right ($\Gamma_+(\lambda)$ and $\Gamma_-(\lambda)$ respectively)
	are related though the jump matrix $G(\lambda)$. 
\end{defn}
\br
The $(-)^{2N-1}$ (which --of course-- is $-1$) is  only reminding that the signs in the first row of $G(\lambda)$ are alternating.
\er

\begin{prop}
	The Riemann-Hilbert problem (\ref{PRH}) is the RH problem associated to the integrable operator 
	\bea
		{\widetilde K(\lambda,\mu)}&:=& \frac {\vec f^{\mathrm{T}}(\lambda) \vec g(\mu)}{\lambda-\mu}\\
		\vec f(\lambda)&:=& \frac{1}{2\pi i}\le(\le[
		\begin{array}{c}  
		{\rm e}^{
		\frac 12 \Theta_0(\lambda)
		}\\
		0\\ \vdots\\ 0
		\end{array} \ri]  \chi_{\gamma_L\cup \gamma_R}(\lambda)+ \le[\begin{array}{c}
		0\\ {\rm e}^{
		-\frac12  \Theta_0(\lambda)+
		 a_1 \lambda} \\ \vdots\\ {\rm e}^{
		 -\frac12  \Theta_0(\lambda) +
		  a_{2N} \lambda}\end{array}\ri]\chi_{i\R}(\lambda)\ri)\\
\vec g(\lambda)&:=& 
\le[\begin{array}{c}
0\\ {\rm e}^{
\frac 12 \Theta_{0}(\lambda)
 -a_1\lambda} \\ \vdots\\ (-)^{2N+1} {\rm e}^{
 \frac 12 \Theta_{0}(\lambda)
  -a_{_{2N}}\lambda} \end{array}\ri]
	 \chi_{\gamma_L\cup\gamma_R}(\lambda)
+
\le[\begin{array}{c}  {\rm e}^{
-\frac 1 2 \Theta_0(\lambda)
}\\
		0\\ \vdots\\ 0
		\end{array} \ri]
 \chi_{i\R}(\lambda)	
\eea
\end{prop}
\noindent{\bf Proof:} It is just enough to verify that $\1 
{-}
2i\pi \vec f(\lambda)\vec g(\lambda)^{\mathrm T}=\1
{-}
 G(\lambda)$. \QED

\begin{theorem} 
\label{KPtilde}
The following identity holds 
	$$\det(\Id-\widetilde K)=\det(\Id-K_{_P}\chi_I)$$
where the operator defined by $\widetilde K$ is of trace-class on $L^2(\gamma_L\cup \gamma_R \cup i\R)$.
\end{theorem}

\noindent{\bf Proof.} 
The proof is essentially identical to that of Thm. \ref{KKtilde} being, in fact, simpler since there  shall be no need of contour deformations.
Writing out $\wt K$ in full we have 
\be
\wt K(\lambda,\mu) = \frac 1{2i\pi} \frac { {\rm e}^{\frac {\Theta_0(\lambda)-\Theta_0(\mu)}2} \chi_{_{i\R }}(\mu)\chi_{_{\gamma_R\cup \gamma_L}}(\lambda)  
- \sum_{j=1}^{2N}  (-)^j {\rm e}^{\frac {\Theta_{0}(\mu) -\Theta_0(\lambda)}2  + a_j \lambda-a_j\mu } \chi_{_{\gamma_R\cup \gamma_L}}(\mu)\chi_{_{i\R}}(\lambda)}{\lambda - \mu}
\ee
Then, using matrix notation subordinated to the split $L^2(\gamma_L\cup \gamma_R) \oplus L^2(i\R)$  we can write $\det(\Id-\widetilde K)$  as
\be
\det\le[\1 - \le[\begin{array}{c|c}
0 & \sum_{j=1}^{2N} (-)^{j+1} \mathcal G_{a_j}\\
\hline
\mathcal F & 0
\end{array}\ri]\ri] = \det[\Id - \sum_{j=1}^{2N} (-1)^{j+1} \mathcal G_{a_j} \circ \mathcal F]
\label{eqB}
\ee
where --this time-- we have defined the operators $\mathcal F,\mathcal G_{a_i}$ as 
\bea
&& \begin{array}{rcl}
 \mathcal G_a: L^2(\gamma_L\cup\gamma_R)&\longrightarrow & L^2(i\R) \\
f(\mu) & \mapsto & \ds \frac{{\rm e}^{-\frac 1 2\Theta_0(\xi)  + a\xi }}{2\pi i}\int_{\gamma_L\cup\gamma_R}{\rm e}^{\frac 12 \Theta_0(\mu) - a\mu} \frac{f(\mu)}{\xi-\mu}\d\mu
\end{array}\\
&&\begin{array}{rcl}
 \mathcal F: L^2(i\R)&\longrightarrow & L^2(\gamma_R\cup\gamma_L)\\
g(\lambda) & \mapsto & \ds \frac{{\rm e}^{\frac 1 2\Theta_0(\mu)} }{2\pi i}\int_{i\R}{\rm e}^{-\frac 12 \Theta_0(\lambda)} \frac{g(\lambda)}{\mu- \lambda}\d\lambda.
\end{array}
\eea
It is a simple verification that the operators $\mathcal G_a, \mathcal F$ are of Hilbert--Schmidt class on $\mathcal H:= L^2(\gamma_L\cup \gamma_R)\oplus L^2(i\R)$; in fact the reader may verify exactly as in Thm. \ref{KKtilde} that both are of trace class by adding a vertical line to the left/right of the imaginary axis  and writing them as composition of HS operators (the check needs to be done separately for $\gamma_L, \gamma_R$). 
Note that the operator $\Id- \sum_{j=1}^{N} (-1)^{j+1} \mathcal G_{a_j} \circ \mathcal F$ appearing in the last equation of (\ref{eqB}) is an operator acting on $L^2(i\R)$ into itself and it is also  of trace-class because resulting from the composition of HS-operators on $\mathcal H$. 
We would like now to conjugate this operator by the (bounded) multiplication operator $\mathcal M:= {\rm e}^{-\frac  1 2 \Theta_0(\xi)}$ on $L^2(i\R)$: this step needs some clarification, since the inverse of this operator is unbounded. By inspection of the operator $\mathcal G_a$ we see that it automatically produces a result in the image of $\mathcal M$ and hence we can multiply it on the left by $\mathcal M^{-1}$. On the other hand, multiplying on the right by $\mathcal M$ causes no problems. The kernel of the resulting operator is thus:
 \be
\mathcal M^{-1} \circ  \mathcal G_{a}\circ \mathcal F\circ \mathcal M  (\xi,\lambda) :=  \frac{1}{(2\pi i)^2} {\rm e}^{ \xi  a }\int_{\gamma_L\cup \gamma_R} \!\!\!\! \d \mu \frac{{\rm e}^{\Theta_0(\mu) -\Theta_0(\lambda) - a\mu }}{(\xi - \mu)(\mu-\lambda)}\label{good}
\ee
Note that one may verify directly that the kernel in (\ref{good}) defines a  trace-class operator on $L^2(i\R)$ by realizing it as the composition of HS operators likewise as above.
The resulting operator is obviously still of trace-class so that the Fredholm determinant is well defined (and has the same value as the original one).
Now we consider the Fourier transform 
\be
& \begin{array}{rcl}
	\mathcal T: L^2(i\R)&\longrightarrow & L^2(\R) \\
f(\mu) & \mapsto & \ds \frac{1}{\sqrt{2i\pi}}\int_{i\R}{\rm e}^{\mu x} f(\mu)\d\mu
\end{array}\quad
\begin{array}{rcl}
	\mathcal T^{-1}: L^2(\R)&\longrightarrow & L^2(i\R) \\
h(x) & \mapsto & \ds \frac{1}{\sqrt{2i\pi}}\int_{i\R}{\rm e}^{-\mu x} h(x)\d x.
\end{array}
\ee
Using (\ref{eqB}) it is enough to prove that $\mathcal T\circ \mathcal M^{-1}\circ(\mathcal G_{a_{2j-1}}-\mathcal G_{a_{2j}})\circ \mathcal F\circ\mathcal M \circ \mathcal T^{-1}= K_P \chi_{[a_{2j-1},a_{2j}]}$. The kernel of the operator $\mathcal T\circ \mathcal M^{-1} \circ \mathcal G_a\circ \mathcal F\circ\mathcal M\circ \mathcal T^{-1}$is equal to  
 \be
  \mathcal L_{a}(x,y) :=  \frac{1}{(2\pi i)^2}\int _{ i\R}\frac{\d\xi}{2\pi i} {\rm e}^{ \xi  (a-x) }\int_{\gamma_L\cup \gamma_R} \!\!\!\! \d \mu \int_{i\R}\!\!\! \d\lambda\,\,\frac{{\rm e}^{\Theta_0(\mu) - \Theta_0(\lambda) - a\mu }}{(\xi - \mu)(\mu-\lambda)}
{\rm e}^{ y \lambda}
 \ee
Now, if $x>a$ we can close the $\xi$--integration with a big circle on the {\em right},  picking up only minus the residue at $\mu\in\gamma_R$; viceversa, if $x<a$ we pick up the residue at $\mu\in \gamma_L$ with the opposite sign; namely 
\bea
 \mathcal L_a(x,y) = \left \{
 \begin{array}{cc}
 \ds  
\mbox{\bf --}\, \frac{1}{(2\pi i)^2}\int_{\gamma_R} \!\!\!\! \d \mu \int_{i \R}\!\!\! \d\lambda\,\,
\frac{{\rm e}^{\Theta_x(\mu)-\Theta_y(\lambda)  }}{(\mu-\lambda)}
 & x>a\\[20pt]
 \ds
 \frac{1}{(2\pi i)^2} \int_{\gamma_L} \!\!\!\! \d \mu \int_{i \R}\!\!\! \d\lambda\,\,\frac{{\rm e}^{\Theta_x(\mu)-\Theta_y(\lambda) }}{(\mu-\lambda)}
  & x<a
  \end{array}
 \right.\label{314}
\eea
Formula (\ref{314}) should be compared with formula (\ref{212}).
Thus we have 
\be
\mathcal L_{a_{2j-1}}(x,y) - \mathcal L_{a_{2j}}(x,y) = \le\{
\begin{array}{cc}
\ds  \frac{1}{(2\pi i)^2}\int_{\gamma_L\cup\gamma_R} \!\!\!\!  \!\!\!\! \d \mu \int_{i \R}\!\!\! \d\lambda\,\,\frac{{\rm e}^{\Theta_x(\mu)-\Theta_y(\lambda)} }{(\lambda-\mu)}=K_{_P}(x,y)& x\in [a_{2j-1},a_{2j}]\\[20pt]
0 & x\not\in [a_{2j-1},a_{2j}]
\end{array}\ri.
\ee
and this last equation concludes the proof. \QED

From Theorem \ref{detMalgrange}, since we have $H(M)=0$ 
 by the same argument used after the proof of Thm. \ref{KKtilde},  we deduce immediately the following
\begin{theorem}\label{FredholmtauII}
	The Fredholm determinant $\det(\Id-K\chi_I)$, with $I=[a_1,a_2]\cup\ldots\cup[a_{2N-1},a_{2N}]$ is equal  to the isomonodromic tau 	function of the RH problem (\ref{PRH}) (note that all intervals are {\em bounded} in the case of the Pearcey kernel).
\be
\pa_{a_\ell} \ln \det(\Id-K\chi_I) = \omega_M(\partial)&\&:= \int_{\gamma_R\cup \gamma_L}\mathrm{Tr}\Big(\Gamma_-^{-1}(\lambda )\Gamma'_-(\lambda)\Xi_\partial(\lambda)\Big)\frac{\d\lambda}{2\pi i}
\ee
Moreover 
\be
\pa_{a_\ell} \ln \det(\Id-K_{_P}\chi_I) =-\Gamma_{1;\ell+1,\ell+1}. 
\ee
\end{theorem}
{\bf Proof.} The proof is identical to the case of the Airy kernel in Thm. \ref{Fredholmtau} and Prop. \ref{Freddef}, simply by changing the phase function $\Theta_a$ from that of Airy (\ref{Airykernel})  to that of Pearcey (\ref{ThetaP}). \QED

\subsection{The solution for \texorpdfstring{$a=b$}{ab}}

When $N=1$ and $a=a_1=b_1=b$ the Fredholm determinant is identically unity, since the action of the kernel is trivial. The Riemann--Hilbert problem is --however-- not immediately trivial. A closer inspection reveals that the solution can be written {\em in closed form} in terms of Cauchy integrals as follows: if we call $A, B, C$ the three columns of $\Gamma$, the jump on $\gamma:= \gamma_L \cup \gamma_R,\  i\R$ and the asymptotics imply
\bea
B(\lambda) &=& \le[\matrix {0\cr 1 \cr 0}\ri] + \frac 1 {2i\pi}\int_\gamma \frac {{\rm e}^{\Theta_a(\mu)}A(\mu)\d \mu}{\mu-\lambda }\ ,\ \ \ 
C(\lambda) = \le[\matrix {0\cr 0 \cr 1}\ri] - \frac 1 {2i\pi}\int_\gamma \frac {{\rm e}^{\Theta_a(\mu)}A(\mu)\d \mu}{\mu-\lambda}\\[6pt]
A(\lambda) &=& \le[\matrix {1\cr 0 \cr 0}\ri] + \frac 1 {2i\pi}\int_{i\R} \frac {{\rm e}^{-\Theta_a(\xi)}(B(\xi)+C(\xi))\d \xi}{\xi-\lambda}
 = \le[\matrix {1\cr 0 \cr 0}\ri] + \le[\matrix {0\cr 1 \cr 1}\ri]  \frac 1 {2i\pi}\int_{i\R}\frac {{\rm e}^{-\Theta_a(\xi)}\d \xi}{\xi-\lambda} 
\eea
and hence the solution is explicitly written for any $a\in \C$ as 
\be
\Gamma(\lambda ) = \1 + \le[
\begin{array}{ccc}
0 &\ds  \int_\gamma \frac { { \rm e}^{\Theta_a(\mu)}\d \mu}{(\mu-\lambda) 2i\pi}  & 
\ds  \int_\gamma \frac { { \rm e}^{\Theta_a(\mu)}\d \mu}{(\mu-\lambda) 2i\pi} \\
\\
 \ds \int_{i\R} \frac { { \rm e}^{-\Theta_a(\xi)}\d \xi}{(\xi-\lambda ) 2i\pi} & 
  \ds \int_\gamma \frac { { \rm e}^{\Theta_a(\mu)}\int_{i\R} \frac { { \rm e}^{-\Theta_a(\xi)}\d \xi}{(\xi-\mu) 2i\pi}\d \mu}
  {(\mu-\lambda) 2i\pi}  &
  \ds - \int_\gamma \frac { { \rm e}^{\Theta_a(\mu)}\int_{i\R} \frac { { \rm e}^{-\Theta_a(\xi)}\d \xi}{(\xi-\mu) 2i\pi}\d \mu}
  {(\mu-\lambda) 2i\pi}  \\
  \\
\ds \int_{i\R} \frac { { \rm e}^{-\Theta_a(\xi)}\d \xi}{(\xi-\lambda) 2i\pi}   & 
\ds \int_\gamma \frac { { \rm e}^{\Theta_a(\mu)}\int_{i\R} \frac { { \rm e}^{-\Theta_a(\xi)}\d \xi}{(\xi-\mu) 2i\pi}\d \mu}{(\mu-\lambda) 2i\pi}   &
 \ds -\int_\gamma \frac { { \rm e}^{\Theta_a(\mu)}\int_{i\R} \frac { { \rm e}^{\Theta_a(\xi)}\d \xi}{(\xi-\mu) 2i\pi}\d \mu}{(\mu-\lambda) 2i\pi}  
\end{array}
\ri]
\ee
This is consistent with the triviality of the Fredholm determinant (identically one, since the kernel is null), which suggests the solvability in ``trivial form'' of the associated RHP. 
Of course a similar expression holds for an arbitrary union $N\geq 1$ of ``empty'' intervals.

\subsection{A Lax system for the Pearcey Process and  PDEs for the gap probability}
\label{PearceyLax}
Using the Riemann-Hilbert problem (\ref{PRH}) it is possible to deduce a Lax system for the Pearcey process. We will treat the case of a single interval with endpoints $a,b$. Let's introduce the matrix
\bea
	\Psi(\lambda;a,b,\tau)&:=&\Gamma(\lambda;a,b,\tau)e^{T(\lambda;a,b,\tau)}\nonumber\\
	T(a,b,\tau;\lambda)&:=&\frac{1}{3}\mathrm{diag}\Big(\Theta_a(\tau;\lambda)+\Theta_b(\tau;\lambda),-2\Theta_a(\tau;\lambda)+\Theta_b(\tau;\lambda),-2\Theta_b(\tau;\lambda)+\Theta_a(\tau;\lambda)\Big)\nonumber\\
\Theta_a (\tau;\lambda) &\& =\frac {\lambda^4}4 - \frac \tau 2 \lambda^2 - a \lambda. 
\eea

In \cite{ACvM1}\footnote{In order to obtain exactly the same equations it is necessary to rescale $\tau\mapsto\tau/2$} two PDEs were found for the Fredholm determinant; we will show that these equation and a third one can be found in this framework, which therefore shows integrability by providing a Lax system.
We will state the equations to be verified and explain the logic of the verification; the actual computation involves a significant amount of completely straightforward algebra and it is best handled by a machine.
\begin{shaded}
\begin{prop}
\label{propPearceyPDE}
The logarithm of the Fredholm determinant 
	$g(a,b,\tau):=\log\det\le(\Id-K_P\big|_{[a,b]}\ri)$ satisfies the differential equations in $\pa_E:= \pa_a + \pa_b$, $\pa_\tau$, $ \epsilon:= a\pa_a + b\pa_b$
	\be
		\pa_E^4g+6(\pa_E^2g)^2-4\tau\pa_E^2g+12\pa_\tau^2g=0
	\label{PearceyPDE}
\ee
\be
\le(-3 \epsilon - 2\tau \pa_\tau +2 \pa_\tau \pa_E^2 +1 \ri)\pa_E g + 12 (\pa_E^2g)(\pa_\tau\pa_E g)=0
\label{PearceyPDE2}
\ee
%
\bea
\epsilon&\& \le(
12 \pa_\tau g - 2 \pa_E^2 g
\ri)
 + \left( 
 8\pa_\tau^2 g
 +
 4\pa_\tau\pa_E^2 g
-4\pa_E^4 g
  -8\, \left( \pa_E^2 g   \right) ^{2} \right) \tau+
\cr
&\& +
4\pa_E^2 g 
+16\, \left(\pa_E^2g \right) ^{3}
+8\, \left( \pa_E\pa_\tau g  \right)
\pa_E^3 g  +
10
\, \left(\pa_E^3 g \right) ^{2}
+16\, \left( \pa_E^4 g   \right) 
\pa_E^2 g  +
\cr
&\& +\pa_E^6 g
 -16\,\pa_\tau^3 g
   +4\,
 \pa_\tau^2\pa_E^2 g    -24\, \left(
 \pa_E\pa_\tau g \right) ^{2}
-8\, \left( 
\pa_\tau\pa_E^2 g  \right)
 \pa_E^2 g
 -8\pa_\tau g  +
 =0
 \label{PearceyPDE3}
\eea
\end{prop}
The latter equation (\ref{PearceyPDE3}) is new.
\end{shaded}
\br
By taking derivatives of the first two and adding them to the third with appropriate coefficients one obtains a somewhat shorter equation
\be
  \pa_E^6 &\& g-8\pa_\tau g
 +4\tau
\left( 2\pa_\tau^3 g-\pa_E^4 g-2 (\pa_E^2 g) ^{2}
 \right)
+12\varepsilon(\pa_\tau g) 
+\cr &\& +
 16(\pa_E^2 g)^{3}
+
4\pa_\tau^2\pa_E^2 g
-24(\pa_\tau\pa_E g)^{2}
+
16\pa_E^4 g \pa_E^2 g+10(\pa_E^3 g)^{2}=0
\ee
It is apparent that this equation cannot be obtained from the first two because it contains a first-order derivative w.r.t. $\tau$ which is absent in the the other equations.
\er
\br If we had only the two equations (\ref{PearceyPDE}, \ref{PearceyPDE2}) the general solution would depend on a functional parameter. The third equation makes the general solution depend on a {\em finite} number of parameters; this fact 
is a manifestation of the underlying isomonodromic system.
\er

We indicate how to verify these statements, but the actual verification (in particular of the third equation) was performed by a machine (the file is available upon request); first we use the new variables 
\be
E = \frac {a+b}2 \ ,\ \ W:= \frac {a-b}2\ , \ \pa_E = \pa_a + \pa_b \ ,\ \ \pa_W = \pa_a-\pa_b\ ,\ \ \ \epsilon:= E\pa_E + W\pa_W
\ee
and use the same symbols for the function $\Psi$ as a function of the new variables $E, W$.
The matrix $\Psi(E,W,\tau;\lambda)$ solves a RHP with constant jumps and has unit determinant; standard arguments allow to conclude easily  that  (the dependence on the variables $E,W,\tau$ shall be understood throughout for brevity)
\be
\pa_\tau \Psi(\lambda)  = L_\tau(\lambda )  \Psi(\lambda) \ & \ 
\pa_E \Psi(\lambda)  = L_E(\lambda )  \Psi(\lambda)\ \pa_W \Psi(\lambda)  = L_W(\lambda )  \Psi(\lambda)\\
& \Psi(\lambda)'  = A(\lambda )  \Psi(\lambda)
\ee
where the matrices $L_\tau, L_E,L_W$ and $A$ are {\bf polynomials} in $\lambda$ of degrees $2,1,1$ and $3$ respectively and $\prime$  denotes the derivative in $\lambda$.

They can be found in the spirit of the \cite{JMU1} as follows.
Denote by $\Gamma_j$ the coefficient matrices in the (asymptotic) expansion of $\Gamma(\lambda)$ near infinity
\be
\Gamma(\lambda) = \1 + \sum_{j=1}^\infty \frac {\Gamma_j} {\lambda ^j}\label{formal}
\ee
In this section we shall then understand $\wh \Gamma$ as the {\bf formal} series appearing in (\ref{formal}) and each instance of $\wh \Gamma^{-1}$ shall be understood as the inversion of the above formal series in the sense of formal series. 
Since we know that all the matrices $L_E,L_W,L_\tau,A$ are polynomials, they can be found by taking the positive part in powers of $\lambda$ (including the constant) of the following expressions
\be
L_\tau(\lambda) = \wh \Gamma \pa_\tau T \wh \Gamma^{-1}  + \pa_\tau \wh \Gamma \wh \Gamma^{-1}\qquad 
L_E(\lambda) = \wh \Gamma \pa_E T \wh \Gamma^{-1}  + \pa_E \wh \Gamma \wh \Gamma^{-1}\\
L_W(\lambda) = \wh \Gamma \pa_W T \wh \Gamma^{-1}  + \pa_W \wh \Gamma \wh \Gamma^{-1}
\qquad A(\lambda) = \wh \Gamma  T' \wh \Gamma^{-1}  +  \wh \Gamma' \wh \Gamma^{-1}
\ee
The important point to make is that the terms like $\pa_E \wh \Gamma \wh \Gamma^{-1}$ etc. are formal series with only negative powers, hence, for example
\be
L_\tau =  \le(\wh \Gamma \pa_\tau T \wh \Gamma^{-1}\ri)_+\ ,\ \ A = \le(\wh \Gamma  T' \wh \Gamma^{-1}\ri)_+
\ee
where $()_+$ denotes the part with nonnegative powers of $\lambda$. It should be evident that this is a polynomial $\lambda$ but also (and more importantly) a polynomial in the coefficient matrices $\Gamma_j$ of the formal expansion, and involving only the first few (the first $2$ for $L_\tau$ and the first $3$ for $A$).

Consider now the ODE in the spectral variable $\lambda$ rewritten for the matrix $\Gamma$ (and also for its formal expansion $\wh \Gamma$)
\be
\Gamma' + \Gamma\, T' = A \Gamma\ .\label{ODEz}
\ee
This is {\em at the same time} a ODE that $\Gamma(\lambda)$ itself satisfies and also its formal expansion $\wh \Gamma$: in terms of the coefficient matrices $\Gamma_j$ this is an {\em infinite system} of polynomial equations in their entries. The coefficients of the positive powers of $\lambda$ in (\ref{ODEz}) satisfy identically the equation by the definition of $A = \le( \wh \Gamma T' \wh \Gamma^{-1}\ri)_+$. The coefficients of the negative powers yield an infinite set of {\bf polynomial} relations between the matrices $\Gamma_j$.

Consider now the equation w.r.t. one of the parameters $E,W,\tau$ (let's take the example of $E$);
\be
\pa_E \Gamma + \Gamma\, \pa_E T = L_E \Gamma\  \ \ 
\Leftrightarrow \ \ \ \pa_E \Gamma = L_E \Gamma -  \Gamma\, \pa_E T
\label{ODEE}
\ee
A straightforward  computation yields
\be
\pa_E T &\&= \frac \lambda 3{\rm diag}(2,-1,-1) =: \frac \lambda 3 \varkappa_1\\
\frac {\pa_E \Gamma_1}\lambda&\& +\dots + \frac \lambda  3 \le(\1 + \frac {\Gamma_1}\lambda  + \frac {\Gamma_2} {\lambda^2} +\dots  \ri) \varkappa_1  =\le( L_E^{(1)} \lambda + L_E^{(0)}\ri) \le(\1 + \frac {\Gamma_1}\lambda  + \frac {\Gamma_2} {\lambda^2} +\dots  \ri)
\ee
From this we have $L_E^{(1)} = \frac 1 3 \varkappa_1$ and $L_E^{(0)} = \frac 1 3 \le[\Gamma_1,\varkappa_1\ri]$ by looking at the coefficients of $\lambda^1, \lambda^0$. The coefficient $\lambda^{-j}$  yield
\be
\pa_E \Gamma_j =   \frac 1 3 \le[\varkappa_1,\Gamma_{j+1}\ri] + \frac 1 3  \le[\Gamma_1,\varkappa_1\ri] \Gamma_j\ ,\ \ j\geq 1.\label{paE}
\ee
A similar computation for $\pa_\tau$ yields
\be
\frac {\pa_\tau \Gamma_1}\lambda &\&+\dots - \frac {\lambda^2}  6 \le(\1 + \frac {\Gamma_1}\lambda  + \frac {\Gamma_2} {\lambda^2} +\dots  \ri) \varkappa_1  =\le( L_\tau^{(2)} \lambda^2 + L_\tau^{(1)} \lambda + L_\tau^{(0)}\ri) \le(\1 + \frac {\Gamma_1}\lambda  + \frac {\Gamma_2} {\lambda^2} +\dots  \ri)\\
L_\tau &\&= \frac 1 6 \le(- {\varkappa_1}\lambda^2 -   \lambda  \le[\Gamma_1,\varkappa_1\ri] +  \le[\varkappa_1 ,\Gamma_2\ri] - \le[\Gamma_1,\varkappa_1\ri]\Gamma_1\ri)\\
\pa_\tau \Gamma_j &\& = \frac 1 6\le( \le[\Gamma_{j+2}, \varkappa_1\ri] -   \le[\Gamma_1,\varkappa_1\ri] \Gamma_{j+1} + \le(\le[\varkappa_1,\Gamma_2\ri] -\le[\Gamma_1,\varkappa_1\ri]\Gamma_1\ri)\Gamma_j \ri)\label{pat}
\ee
Equations (\ref{paE}, \ref{pat}) allow to express any derivative (of any order by using Leibnitz rule) of the coefficient matrices $\Gamma_j$ in terms of {\em polynomials} in the same coefficient matrices (the expressions become larger and larger but stay finite at each step).
In order to verify Prop. \ref{propPearceyPDE} (and eqs. (\ref{PearceyPDE},\ref{PearceyPDE2})) it suffices to write 
\be
\pa_E g &\& = -\res_{\lambda=\infty} \Tr \le(\Gamma^{-1} \Gamma' \pa_E T\ri) = \frac 1 3 \Tr(\Gamma_1 \varkappa_1)\ ,\\
\pa_\tau g &\& = -\res_{\lambda=\infty} \Tr \le(\Gamma^{-1} \Gamma' \pa_\tau T\ri) =  \frac 1 6 \Tr\le((\Gamma_1^2 - 2\Gamma_2) \varkappa_1\ri)\ ,
\ee
One then makes repeated use of eqs. (\ref{paE}, \ref{pat}) plugging the result into equations (\ref{PearceyPDE},\ref{PearceyPDE2}); the result is a rather large polynomial in the coefficients matrices $\Gamma_1, \Gamma_2, \Gamma_3$.

One then compares with the following polynomials 
\be
\res_{\lambda=\infty}\Tr \le(\le( \Gamma' \Gamma^{-1}+ \Gamma \, T'\Gamma^{-1} - A\ri) \pa_E T\ri)\frac{\d \lambda}{\lambda^j}\ ,\ \ \ j=1,2,3
\ee
which have to be identically zero due to (\ref{ODEz}) and verifies that they are the same polynomial thus completing the check.
\section{Large asymptotics for the Pearcey Gap probability and factorized Tracy--Widom asymptotic behavior}

The goal of this section is to prove the asymptotic factorization of the Fredholm determinant for a ``large'' gap in the Pearcey process into two Fredholm determinants for semi--infinite gaps of the Airy process. In detail 
\begin{shaded}
\begin{theorem}
\label{main1}
Let $K_{_P}(x,y;\tau)$ denote the Pearcey kernel and $K_{\Ai}(x,y)$ the Airy kernel. 
Let 
\be
a_-=a_-(\rho):= -2\Lambda^9 + \Lambda \rho \sqrt[3]{3}\ \ \ \ 
a_+=a_+(\sigma):= 2\Lambda^9 - \Lambda \sigma \sqrt[3]{3}\ \ \ 
\tau:= 3\Lambda^6
\ee
then as $\Lambda\to+\infty$ 
\be
\det\le(\Id - K_{_P}(\bullet,\bullet;\tau)\bigg|_{[a_-,a_+]}\ri) =\det\le(\Id - K_\Ai(\bullet,\bullet)\bigg|_{[\rho,\infty)}\ri) \det\le(\Id - K_\Ai(\bullet,\bullet)\bigg|_{[\sigma,\infty)}\ri) (1 + \mathcal O(\Lambda^{-1}))\nonumber
\ee
and the convergence is uniform over compact sets of the variables $\rho,\sigma$ of the form 
\be
K_1 \leq \rho,\sigma \leq \frac {K_2}{\sqrt[3]{3}} \Lambda^8,\ \ -\infty<K_1, \ K_2<2,
\ee 
namely for which  $\rho,\sigma$ are uniformly bounded below (w.r.t. $\Lambda$) and do not grow faster than $\tau^{\frac 43}$.
\end{theorem}
\end{shaded}

\begin{figure}[h]
\resizebox{0.5\textwidth}{!}{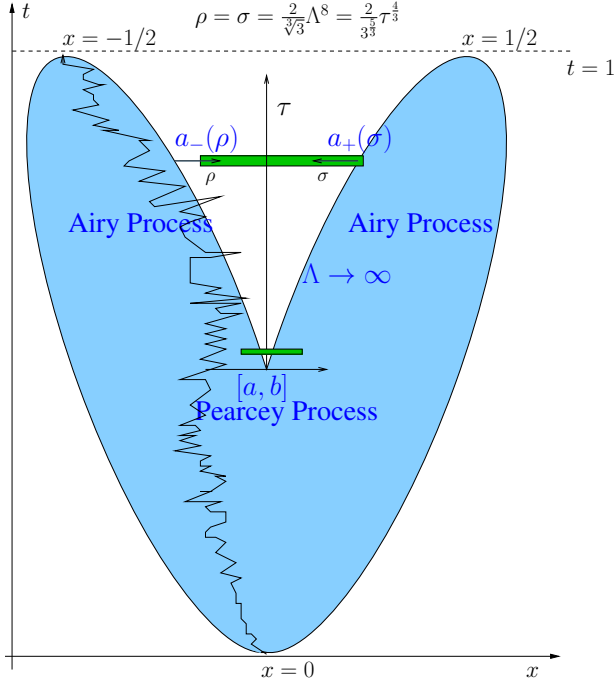}
\caption{The Pearcey and Airy processes.}
\label{FigPtoA}
\end{figure}

More generally  when we have  a union of intervals;
\begin{theorem}
\label{main2}
Let $\mathfrak I$ be a union of intervals of the form 
\bea
\mathfrak I &\&= \bigcup_{j=1}^J [a_{2j-1}, a_{2j}] \cup[a_{2J+1}, b_{0}] \cup \bigcup_{k=1}^{K} [b_{2k-1}, b_{2k}]\\
a_\ell&\& :=a_-(\rho_\ell) =  -2\Lambda^9 + \Lambda \rho_\ell \sqrt[3]{3}\qquad 
b_{\ell} :=a_+(\sigma_{2K+1-\ell}) = 2\Lambda^9 - \Lambda \sigma_{2K+1-\ell} \sqrt[3]{3}\ ,\ \ \ \tau = 3\Lambda^6
\ee 
Then, as $\Lambda\to \infty$ we have 
\bea
\det\le(\Id - K_{_P}(\bullet,\bullet;\tau)\bigg|_{\mathfrak I}\ri) =\det\le(\Id - K_\Ai(\bullet,\bullet)\bigg|_{J_1}\ri) \det\le(\Id - K_\Ai(\bullet,\bullet)\bigg|_{J_2}\ri) (1 + \mathcal O(\Lambda^{-1}))\\
J_1:= \bigcup_{\ell=1}^J [\rho_{_{2\ell -1}}, \rho_{_{2\ell}}]\cup [\rho_{_{2J+1}},\infty)\qquad
J_2 :=\bigcup_{\ell=1}^K [\sigma_{_{2\ell -1}}, \sigma_{_{2\ell}}]\cup [\sigma_{_{2K+1}},\infty) 
\eea
In the above it is allowed that $a_{2J+1} = b_0$ (hence the middle interval is missing) in which case $J_1:= \bigcup_{\ell=1}^J [\rho_{_{2\ell -1}}, \rho_{_{2\ell}}]$ and $J_2 :=\bigcup_{\ell=1}^K [\sigma_{_{2\ell -1}}, \sigma_{_{2\ell}}]$.
\end{theorem}

The proof of this theorem consists of the whole present section; it relies essentially upon the Riemann--Hilbert problem constructed in Sect. \ref{PearceyKernel} and the Deift--Zhou \cite{DeiftZhou} steepest descent method.
We will start with considerations that apply to the more general case of Theorem \ref{main2} but then specialize to the case of Theorem \ref{main1} in order to avoid unnecessary complications (which are purely notational and not conceptual).

\begin{remark}
The parametrization of the endpoints $a_-, a_+$ in Thm. \ref{main1} (and of $a_j, b_j$ in Thm. \ref{main2}) has the following meaning.
The Pearcey process arises in the study of  {\em self--avoiding random walks}  on the line, conditioned to start at the same point (say the origin) at time $t=0$ and end at time $t=t_1>0$ at two distinct points moving away from the origin as $N^{1/2}$ (N being the number of particles); at any time $0<t<t_1$ the bulk of the walkers consists of either one or two finite intervals. There is a critical time $0<t_c<t_1$ at which this bulk undergoes a transition from a connected interval to two intervals. The two new emerging endpoints $[a(t),b(t)]$  move away from a common point $a(t_c)=b(t_c)$ according to $a(t)= a_c + \mathcal O(t-t_c)^{\frac 32}$.  The Pearcey point process describes the statistics of the random walkers in a scaling neighborhood of $t=t_c$ and $a,b=a_c$; more precisely we rescale $t\mapsto t_c+N^{-1/2}t$, $a\mapsto a_c+N^{-1/4}a$ and $b\mapsto a_c+N^{-1/4}b$; see for instance \cite{TracyWidomPearcey} . The asymptotics as $\tau=3\Lambda^6\to \infty$ and $a=a_-,b=a_+$ as given in Thm. \ref{main1}, is the regime where we look ``away'' from the critical point and it is expected to reduce to two Airy point processes, which describe the edge-behavior of the random walkers.
\end{remark}

The phase $\Theta_a(\tau;\lambda)$ (\ref{ThetaP}) has an inflection point with zero derivative when the discriminant of the derivative vanishes:
\be
{\rm Discrim}_\lambda(\lambda^3 - \tau\lambda - a) = -4 \tau^3 + 27 a^2 = 0
\ee
The neighborhood of the discriminant is  parametrizable as follows:
\be
a_\pm(s) := \pm (2 \Lambda^9 -  \Lambda s\sqrt[3]{3})\ ,\ \ \ \tau = 3 \Lambda^6\label{apm}
\ee
We have the expression
\bea
\Theta_{a_\pm}(\tau;\lambda) &\& = \overbrace{\frac 34 \Lambda^{12} - 3^{\frac 13  } \Lambda^4 s}^{=:C(s)} \mp \overbrace{\le(\frac 13  \zeta^3 - s\zeta\ri)}^{=: \vartheta_s(\zeta)} + \frac {(\lambda\pm\Lambda^3)^4}4=\\
&\& = {C(s)} \mp {\vartheta_s(\zeta)} +  \frac {(\lambda\pm\Lambda^3)^4}4\label{Cdef}\\
&\& \zeta:=\zeta_{L,R} =  3^{\frac 1 3}  \Lambda (\lambda\pm  \Lambda^3)
\label{defzetaRL}
\eea
For definiteness we consider only the case $a_{2J+1} < b_0$.

{\bf Preliminary step}: we conjugate $\Gamma$ by the diagonal constant matrix $D_0$ with entries
\bea
(D_0)_{11}:= \frac 1{2N+1} \le( \sum_{\ell=1}^{2J+1} C(\rho_\ell) + \sum_{\ell=1}^{2K+1} C(\sigma_\ell)\ri) \\
(D_0)_{\ell+1,\ell+1} := (D_0)_{11} - C(\rho_{\ell})\ ,\ \ \ \ell=1\dots 2K+1\\
(D_0)_{\ell+2J+2, \ell+2J+2} = (D_0)_{11} - C(\sigma_{\ell})\ ,\ \ \ \ell=1\dots 2J+1.
\eea
with $C(s) = \frac 34 \Lambda^{12} - 3^\frac 13 \Lambda^4 s$  as in (\ref{Cdef}). 
We denote by a tilde the new matrix and respective jumps
\be
\wt \Gamma(\lambda) := {\rm e}^{-D_0} \Gamma(\lambda) {\rm e}^{D_0} \ ,\qquad 
\wt M(\lambda):= {\rm e}^{-D_0} M(\lambda) {\rm e}^{D_0}\ .
\ee
The conjugation by $D_0$ has simply the effect of replacing the phases $\Theta_{a_-(\rho_\ell)},\Theta_{a_-(\sigma_\ell)}$ by $\Theta_{a_-(\rho_\ell)}- C(\rho_\ell)$ and $\Theta_{a_+(\sigma_\ell)}-C(\sigma_\ell)$, respectively,  so that their critical value is zero.

{\bf Factorization of the jumps}: the jump on the imaginary axis $\wt M_0:= \wt M\big|_{i\R}$ for  $\wt \Gamma$ can be factorized 
\bea
&\& \wt M_0 = \wt M_{0,R}\wt M_{0,L}^{-1}\ ,\qquad \wt \Theta_{a_\pm (s)} (\lambda):= \Theta_{a_\pm(s)} (\lambda)  - C(s)\\
&\& \wt M_{0,L} :=
\!\!\! \le[ \begin{array}{c|c|c}
		1&&\\\hline
		\vdots&\1_{2J+1}&\\
		0 &&\\
		\hline
		{\rm e}^{-\wt \Theta_{a_{+}(\sigma_{_{2K+1}})}}
		&&\\
		\vdots&&\1_{2K+1}\\
		{\rm e}^{-\wt \Theta_{a_{+}(\sigma_{_{1}})}}&&
		\end{array}\ri]
		\ 
\wt M_{0,R} :=\!\!\! \le[ \begin{array}{c|c|c}
		1&&\\\hline
		{-}
		{\rm e}^{-\wt \Theta_{a_{-}(\rho_{_1})}} &&\\
		\vdots&\1_{2J+1}&\\
		{-}
		{\rm e}^{-\wt \Theta_{a_{-}(\rho_{_{2J+1}})}} &&\\
		\hline
		\vdots&&\1_{2K+1}\\
		0&&
		\end{array}\ri]
		\label{factorization}
\eea
This factorization  allows us eventually  to replace the jump on $i\R$ by a jump on two separate contours in the right/left halfplanes; this will be used in the final step below.

{\bf Rescaling}: in view of the fact that the critical points $\lambda  = \pm \Lambda^3$ escape to infinity, we now introduce the scaled variable $z:= \Lambda^{-3} \lambda$.
\begin{figure}
\resizebox{0.7\textwidth}{!}{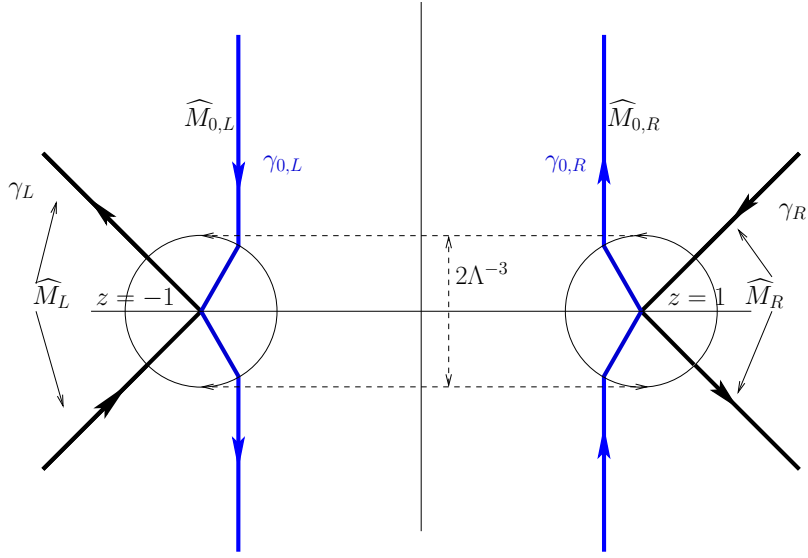}
\caption{The contours of jumps for the RHP for  $ Y$.}
\label{FigModRHP}
\end{figure}

{\bf Final step:}
We shall choose the contours for our final matrix as follows in the plane $z = \Lambda^{-3} \lambda$:
\begin{itemize}
\item Two counterclockwise circles centered at $z = \pm 1$ of radii $r = \Lambda^{-3}$ (i.e. of radius $1$ in the $\lambda$--plane);
\item the two contours $\gamma_{R,L}$ shall be moved to the contours consisting of the two pairs of straight  half-lines originating a $z = \pm 1$ ($\lambda=\pm \Lambda^3$) with slopes $\pm \frac \pi 4$;
\item the two contours that shall support the jump that was originally on $i\R$ will be denoted by $\gamma_{0,R}$ and $\gamma_{0,L} = -\gamma_{0,R}$. The contour $\gamma_{0,R}$ is 
\be
\gamma_{0,R}:= \{\Re z = 1 - \Lambda^{-3} {\rm e}^{\frac{2i\pi}3},\ |z-1|\geq 1\}  \cup
 \{ 1 + s{\rm e}^{\pm \frac {2i\pi}3}  \ ,\ s\in [0,\Lambda^{-3}) \}, 
\ee
and oriented upwards.
\end {itemize}
The arrangement of these contours is best illustrated in Fig. \ref{FigModRHP}. Define now 
\be
Y(z):=\le\{
\begin{array}{cc}
\wt  \Gamma( \Lambda^3 z) \wt M_{0,R}(\Lambda^3 z)  & \hbox{in the region between $i\R$ and $\gamma_{0,R}$}\\
\wt  \Gamma( \Lambda^3 z) \wt M_{0,L}(\Lambda^3 z)  & \hbox{in the region between $i\R$ and $\gamma_{0,R}$}\\
\wt \Gamma(\Lambda^3 z) & \hbox{ elsewhere.}
\end{array}
\ri.
\ee
The matrix $Y(z)$ solves a new RHP ($\wh M(z) = \wt M(\Lambda^3 z)$) with jumps:
\bea
Y_+ &\& = Y_- \wh M\ ,\qquad Y(z)=\1 + \mathcal O(z^{-1})\ ,\cr 
\wh M(z)&\& := \wh M_{0,R}(z) \chi_{\gamma_{0,R}} + \wh M_{0,L}( z) \chi_{\gamma_{0,L}} + \wh M_{R}(z) \chi_{\gamma_{R}} + \wh M_{L}( z) \chi_{\gamma_{L}}
\label{YRHP}
\eea
\be
&\& \wh M_{L}=\wh M_{R} :=
 \le[\begin{array}{c|c|c}
		1 &{-}
		 {\rm e}^{\wh \Theta_{a_{-}(\rho_1)}}
		 \dots  (-)^{2J{+1}}
		   {\rm e}^{\wh  \Theta_{a_{-}(\rho_{_{2J+1}})}} &
		   {\rm e}^{\wh \Theta_{a_{+}(\sigma_{_{2K+1}})}} 
		   \dots  (-)^{2N } {\rm e}^{\wh \Theta_{a_{+}(\sigma_{_{1}})}}  \\
		\hline
		& &\\ 
		&\hspace{20pt} \1_{2J+1}&\\
		\hline
		&&\1_{2K+1}\\
		&&
\end{array}\ri]\label{415}\\
&\& \wh M_{0,L} :=
\!\!\! \le[ \begin{array}{c|c|c}
		1&&\\\hline
		\vdots&\1_{2J+1}&\\
		0 &&\\
		\hline
		{\rm e}^{-\wh \Theta_{a_{+}(\sigma_{_{2K+1}})}}
		&&\\
		\vdots&&\1_{2K+1}\\
		{\rm e}^{-\wh \Theta_{a_{+}(\sigma_{_{1}})}}&&
		\end{array}\ri]
		\ 
\wh M_{0,R} :=\!\!\! \le[ \begin{array}{c|c|c}
		1&&\\\hline
		{-}
		{\rm e}^{-\wh \Theta_{a_{-}(\rho_{_1})}} &&\\
		\vdots&\1_{2J+1}&\\
		{-}{\rm e}^{-\wh \Theta_{a_{-}(\rho_{_{2J+1}})}} &&\\
		\hline
		\vdots&&\1_{2K+1}\\
		0&&
		\end{array}\ri]\\
&\& \wh \Theta_{a_\pm(s)}(z):= \Theta_{a_{\pm}(s)} ( 3\Lambda^6; \Lambda^3 z) - C(s)= \mp \vartheta_s (\zeta) + \Lambda^{12} \frac {(z\pm 1)^4} 4\\
&\& \zeta := \zeta_{L, R} =  \sqrt[3]{3} \Lambda^4 (z\pm 1)\ ,\ \ \  \vartheta_s(\zeta):= \frac {\zeta^3}3 - s \zeta.
\eea
(The subscripts $L,R$ in $\zeta$ mean that one is centered around the {\bf L}eft critical point $z = -1$ and the other around the {\bf R}ight critical point $z = 1$). The two circles for the time being do not support any jump of $Y$, but will be used eventually in the construction of the approximation.

The main idea of the proof is now that 
\begin{itemize}
\item The matrix $\wh M_{0,R}$ 
  is exponentially close to the identity in $L^p(\gamma_{0,R})$ $p\leq \infty$ outside of the disk at $z=1$. Similarly for $\wh M_{0,L}$ relative to $\gamma_{0,L}$. 
\item For the  matrix $\wh M_R$,  the entries of the form ${\rm e}^{\wh \Theta_{a_{+}(\sigma_\ell)}}$ are all uniformly small (as $\Lambda\to\infty$) in any $L^p(\gamma_R)$, $p\leq \infty$; ditto for $\wh M_L$ and the entries ${\rm e}^{\wh \Theta_{a_{-}(\rho_\ell)}}$. 
\item Inside the disks we will use an parametrix build out of the AiORHP (\ref{multiAi}) of the appropriate size that solves the jumps up to exponentially and uniformly small terms.
\end{itemize}
The proofs of the first two bullet points rely upon Lemma \ref{Lemma1},  Lemma \ref{Lemma2} and Lemma \ref{Lemma3}.
\begin{lemma}
\label{Lemma1}
Let $K_2<2$ be fixed and $\sigma\leq K_2 3^{-\frac 1 3} \Lambda^8$; then the function ${\rm e}^{\wh \Theta_{a_{+}(\sigma)}(z)}$  ($\tau = 3\Lambda^6$) tends to zero exponentially (in $\Lambda$) in any norm $L^p(\gamma_R)$,  $1\leq p \leq \infty$
\be
\|
{\rm e}^{\wh \Theta_{a_{+}(\sigma )}(z)} \|_{L^p(\gamma_R)} \leq  {\rm e}^{-2(2-K_2)\Lambda^{12}}\ .
\ee
 Similarly, the function ${\rm e}^{\wh \Theta_{a_{-}(\rho)}(z)}$   tends to zero exponentially (in $\Lambda$) in any norm $L^p(\gamma_L)$,  $1\leq p \leq \infty$.
\end{lemma}
{\bf Proof.}
We set $z = 1  + (1\pm i) t$, $t\in \R_+$ as a parametrization of $\gamma_R$. A trivial computation yields (for both signs!)
\be
H(t):= \Re(\wh \Theta_{a_{+}(\sigma)}(z)) =   \Lambda^{12}
 \le(
 -t^4 - 2 t^3 + t\le( \frac {\sigma \sqrt[3]{3}} {\Lambda^8} - 4  \ri) \ri) + 2\Lambda^4 \le(\sigma \sqrt[3]{3}  - 2\Lambda^8\ri)
\ee
We have $|{\rm e}^{\wh \Theta_{a_{+}(\sigma)}}| = {\rm e}^{H(t)}$: under the assumption that $\sigma \leq K_2 3^{-\frac 1 3} \Lambda^8$ with a constant $K_2<2$ (independent of $\Lambda$) we have (recall that $t\in \R_+$)
\bea
H(t)\leq -\Lambda^{12} t^4 - 2\Lambda^{12} (2-K_2)\\
\| {\rm e}^H\|_{L^p(\gamma_R)}\leq  {\rm e}^{-2(2-K_2)\Lambda^{12}} \le(2\int_0^\infty {\rm e}^{-p\,\Lambda^{12} t^4}\d t\ri)^{\frac 1 p} \leq  {\rm e}^{-2(2-K_2)\Lambda^{12}} \le(\frac {\pi }{\sqrt{2} p^{\frac 1 4}\Lambda^3  \Gamma(3/4)}\ri)^{\frac 1 p}
\eea
One verifies that this last expression of $p$ is less than $1$ (for $\Lambda>1$).
Note that if $K_2=2$ and $\sigma$ saturates the upper bound, one can still show that the $L^p$ norms are exponentially small, except the $L^\infty$ which is constant.
The other case is completely analogous.
\QED
\br
In Lemma \ref{Lemma1} the upper bound $K_2=2, \sigma=K_2 3^{-1/3} \Lambda^8$ corresponds to $a_+(\sigma) =0$, so the above estimate holds as long as $a_+>0$ and $\tau\to \infty$ (see Fig. \ref{FigPtoA}).
\er
\begin{lemma}
\label{Lemma2}
Let $K_1>-  \infty$ be arbitrarily fixed; then, uniformly in $\rho \geq K_1$ 
the function ${\rm e}^{\wh \Theta_{a_{-}(\rho)}(z)}$ tends to zero exponentially (in $\Lambda$) in any norm $L^p(\gamma_{R,e})$,  $1\leq p \leq  \infty$, where $\gamma_{R,e}:= \gamma_R \setminus \{|z-1|\leq \Lambda^{-3}\}$ and specifically 
\be
\|{\rm e}^{\wh \Theta_{a_-(\rho)}(z)}\|_{L^p(\gamma_{R,e})} \leq  {\rm e}^{ - \frac {\sqrt{2}}{2} \Lambda^3  + \frac {\sqrt{2}}2  \Lambda|K_1| \sqrt[3]{3}}
\ee
Similarly for the function ${\rm e}^{\wh \Theta_{a_{+}(\sigma)}(z)}$ in  $L^p(\gamma_{L,e})$,  $\gamma_{L,e}:= \gamma_L \setminus \{|z+1|\leq \Lambda^{-3}\}$.
\end{lemma}
{\bf Proof.}
The rays of $\gamma_{R,e}$ can be parametrized as 
We set $z = 1  +{\rm e}^{\pm \frac {i\pi}4}  (t+\Lambda^{-3})$, $t\geq 0$. A trivial computation yields (for both signs!)
\be
H(t):= \Re(\wh \Theta_{a_{-}(\rho)}(z))  =
 \Lambda^{12}\le(-\frac 1 4\,{t}^{4}-{\frac { \left( \sqrt {2}{{\Lambda}}^{3}+2
 \right) {t}^{3}}{{{2\Lambda}}^{3}}}-\frac 3 2{\frac { \left( \sqrt {2}{
{\Lambda}}^{3}+1 \right) {t}^{2}}{{{\Lambda}}^{6}}}+\ri.\\
\le.+ \left( -
\,{\frac {\sqrt {2}{\rho}\,\sqrt [3]{3}}{{{2\Lambda}}^{8}}}-
{\frac {2+3\,\sqrt {2}{{\Lambda}}^{3}}{{{2\Lambda}}^{9}}}
 \right) t-{\frac {\sqrt {2}{\rho}\,\sqrt [3]{3}}{{{2\Lambda
}}^{11}}}-{\frac {1+2\,\sqrt {2}{{\Lambda}}^{3}}{{{4\Lambda}
}^{12}}}\ri)
\ee
The coefficient of the linear term in $t$ is negative for $\rho>K_1$ and  sufficiently large $\Lambda$ hence we can estimate
\be
H(t) \leq - \Lambda^{12} \frac {t^4}4  - \frac {\sqrt{2}}{2} \Lambda^3  + \frac {\sqrt{2}}2  \Lambda|K_1| \sqrt[3]{3}.
\ee
 This shows that the $L^\infty$ norm tends to zero since $H(t)\leq H(0) \leq - \frac {\sqrt{2}}{2} \Lambda^3  + \frac {\sqrt{2}}2  \Lambda|K_1| \sqrt[3]{3}\to -\infty$. As for the other $L^p$ norms, we have
\be
\|{\rm e}^H\|_{L^p(\gamma_{R,e})}^p = \int_{\Lambda^{-3}}^\infty {\rm e}^{pH(t)} \d t \leq {\rm e}^{pH(0)}  \int_{\Lambda^{-3}}^\infty {\rm e}^{p(H(t)-H(0))} \d t = {\rm e}^{pH(0)}  \int_{0}^\infty {\rm e}^{-\frac {p\Lambda^{12}}{4} t^4} \d t ={\rm e}^{pH(0)} \le(\frac {\pi \sqrt{2}} 
{4^\frac 34 p^\frac 14 \Lambda^3 \Gamma(\frac 34)
}\ri)\nonumber 
\ee
Hence (for sufficiently large $\Lambda$ the last bracket is smaller than one).
\be
\|{\rm e}^H\|_{L^p(\gamma_{R,e})} \leq {\rm e}^{H(0)} \leq {\rm e}^{ - \frac {\sqrt{2}}{2} \Lambda^3  + \frac {\sqrt{2}}2  \Lambda|K_1| \sqrt[3]{3}}\to 0
\ee
\QED
\begin{lemma}
\label{Lemma3} Let $K_1 >-\infty$ be arbitrarily fixed; then, uniformly in $\rho\geq K_1 $ the function ${\rm e}^{-\wh \Theta_{a_{-}(\rho)}(z)}$  tends to zero exponentially (in $\Lambda$) in any $L^p$, $1\leq p \leq\infty$, of the vertical rays of $\gamma_{0,R}$, denoted by $\gamma_{0,R,e}$,  in particular
\be
\|{\rm e}^{-\wh \Theta_{a_-(\rho)}(z)}\|_{L^p(\gamma_{0,R,e})} \leq  {\rm e}^{-\Lambda^3 +  K_1 \Lambda/2 +1}\ ,\ \ \ \gamma_{0,R,e}:= \gamma_{0,R} \setminus \{|z-1|<\Lambda^{-3} \}
\ee
 Similarly for ${\rm e}^{-\wh \Theta_{a_{+}(\sigma)}(\tau;\lambda)}$ on the vertical rays  $\gamma_{0,L,e}:= \gamma_{0,L} \setminus \{|z+1|<\Lambda^{-3} \}$.
 \end{lemma}
{\bf Proof.}
Once more this is a trivial check using $z = 1+ \frac {{\rm e}^{\pm 2 i\pi/3}}{\Lambda^3} \pm  i t$, $t\in \R_+$ 
\bea
H(t)&\& := - \Re(\wh \Theta_{a_{-}(\rho)}(z)) =\\&\&=\Lambda^{12}\le(
 -\frac 1 4\,{t}^{4}-\frac 1 {2\Lambda^3} \,\sqrt {3}{ t}^{3}- \left( \frac 3 {2\Lambda^{3}}+\frac 3 
{4\Lambda^6} \right) {t}^{2}-\frac {3\sqrt[3]{3}} {2\Lambda^{6}} t\ri) +\frac 1 8-\frac 1 2\,{\Lambda}\,\rho \,\sqrt [3]{3}-{{\Lambda}}^{3}
\\
&\& \leq
 -\frac {\Lambda^{12}} 4 t^4  - \Lambda^3 - \frac {\sqrt[3]{3}}2\Lambda\rho  + \frac 18 \leq  -\frac {\Lambda^{12}} 4 t^4  - \Lambda^3 -  K_1 \Lambda/2 + 1
\ee
Thus we have 
\bea
\|{\rm e}^{-\wh \Theta_{a_-(\rho)}(z)}\|_{L^p(\gamma_{0,R,\pm})} = \le(2\int_0^\infty {\rm e}^{p H(t)} \d t \ri)^\frac 1 p \leq{\rm e}^{-\Lambda^3 - \frac{K_1}2  \Lambda +1 }
 \le(\frac {\pi }{2 p^{\frac 1 4}\Lambda^3  \Gamma(3/4)}\ri)^{\frac 1 p}
\eea
Hence it is clear that ${\rm e}^{H(t)}$ satisfies the assertion (for large $\Lambda$, e.g. $\Lambda>4$)  because the constant in the $p$--th root is less than one for  $\Lambda$ large. Completely analogous computation on $\gamma_{0,L}$.
\QED

In order to obtain the desired asymptotic information one needs to solve the RHP within the two disks centered at $z=\pm 1$; these local RHP are expressed in terms of $\zeta_{R,L}$ (\ref{defzetaRL}); the local parametrices are then constructed in terms of the RHPs for the Airy kernel in the form  (\ref{AIRH}).  

As we have anticipated, we will enter into details now for the simplest case where $K=J=0$ and hence we have a single interval for the Pearcey kernel of the form 
$
[a_{-}(\rho), a_{+}(\sigma)]
$.
The jumps for $Y$ are explicitly 
\be
Y(\lambda)_+&\&  =Y_-(\lambda) \wh M_{0,R}\ \ \ \ \ \lambda\in \gamma_{0,R}\ ;\qquad \ Y(\lambda)_+  =Y_-(\lambda) \wh M_{0,L}\ \ \ \ \ \lambda\in \gamma_{0,L}  \\
\wh M_{0,L}&\& :=    \le[
\begin{array}{ccc}
1 & 0 & 0 \\
0& 1 & 0\\
{\rm e}^{ \vartheta_\sigma (\zeta_L) - \frac 14 \lambda_L^4}&0&1
\end{array}
\ri]\ ,\qquad
\wh M_{0,R}:=\le[
\begin{array}{ccc}
1 & 0 & 0 \\
{-}
{\rm e}^{ -\vartheta_\rho (\zeta_R)- \frac 14 \lambda_R^4}& 1 & 0\\
0&0&1
\end{array}
\ri] \\
Y(\lambda)_+ &\& = Y_-(\lambda) \wh M_L\ \ \ \ \lambda\in \gamma_L\qquad
Y(\lambda)_+  = Y_-(\lambda) \wh M_R\ \ \ \ \lambda\in \gamma_R
\\
\wh M_R=\wh M_L&\& :=\le[
\begin{array}{ccc}
1 &   {-}
{\rm e}^{\vartheta_\rho(\zeta_R) + \frac 14 \lambda_R^4} &
{\rm e}^{ -\vartheta_\sigma (\zeta_L) + \frac 14 \lambda_L^4}\\
0& 1 & 0\\
0&0&1
\end{array}
\ri] \\
&& \lambda_R:= (\lambda-\Lambda^3) = \Lambda^3(z-1) \ ,\qquad \lambda_L:= (\lambda+\Lambda^3) = \Lambda^{3}(z+1)\label{llr}\ .
\ee
Note that  the matrices $\wh M_{R}, \wh M_L$ have identical form, but are defined respectively on $\gamma_R, \gamma_L$. 
Also important is the fact that the terms $\lambda_{L}^4, \lambda_R^4$ (with $\lambda_L, \lambda_R$ defined in (\ref{llr})) appearing in the exponents are uniformly bounded within the disks because these have radius $\Lambda^{-3}$ in the $z$--plane (or radius $1$ in the $\lambda$--plane).
As a result of Lemmas \ref{Lemma1}, \ref{Lemma2}, \ref{Lemma3}, the jumps on the contours outside the two discs at $\pm \Lambda^3$ are exponentially close to the identity matrix in any $L^p$ norm (including $L^\infty$). 

Moreover, thanks to Lemma \ref{Lemma1},  the jumps on $\gamma_R$ and $\gamma_L$  can be written as 
\bea
\wh M_{R}(z) &\&=\le(\1 +\mathcal O\le( {\rm e}^{-2(2-K_2) \Lambda^{12}}\ri)\ri)\wh M^{(0)}_R(z) \ ,\ \ 
 \wh M_R^{(0)}:= \le[
\begin{array}{ccc}
1&  
{-}
{\rm e}^{\vartheta_\rho(\zeta_R)+ \frac 14 \lambda_R^4}  &0\\
0& 1 & 0\\
0&0&1
\end{array}
\ri]
\label{538}\\
\nonumber\\
\wh M_{L}(z) &\&=\le(\1 + \mathcal O\le( {\rm e}^{-2(2-K_2) \Lambda^{12}} \ri)\ri)\wh M^{(0)}_L\ ,\ \ 
  \wh M_L^{(0)} :=  \le[
\begin{array}{ccc}
1&  0  &
{\rm e}^{ -\vartheta_\sigma (\zeta_L) + \frac 14 \lambda_L^4}\\
0& 1 & 0\\
0&0&1
\end{array}
\ri]
\label{539}
\eea
where the estimate holds
uniformly on $\rho, \sigma < K_2 3^{-\frac 1 3} \Lambda^8$ for any $K_2<2$.

\subsection{Approximation of the solution}

{\bf Notation}: Given a $2\times 2$ matrix $A$ we will define the $3\times 3$ matrix $A^{(i,j)}$, $1\leq i\neq j \leq 3$ as the matrix whose elements $(i,i), (i,j), (j,i)$ and $(j,j)$ are the elements $(1,1), (1,2), (2,1), (2,2)$ of $A$ and zero elsewhere.
\subsubsection{Parametrix near \texorpdfstring{$z=1$}{z1}}
In this subsection and in the following we use the Hasting-McLeod  matrix $\Gamma_{\mathrm{Ai}}$ (defined in Section \ref{auxiliarymatrix}  and whose asymptotics is discussed ibidem, and which is nothing but the solution of the RHP in Def. \ref{multiAi} for $N=1$) as parametrix for the RH problem related to $Y$ inside the disks centered at $z=\pm 1$. 
The same parametrix was used in several papers in different contexts, e.g. \cite{BaikDeiftJohansson99, BleherIts}.
Let's start with the disk centered in $z=1$. Consider the  matrix 

\begin{figure}
\resizebox{0.4\textwidth}{!}{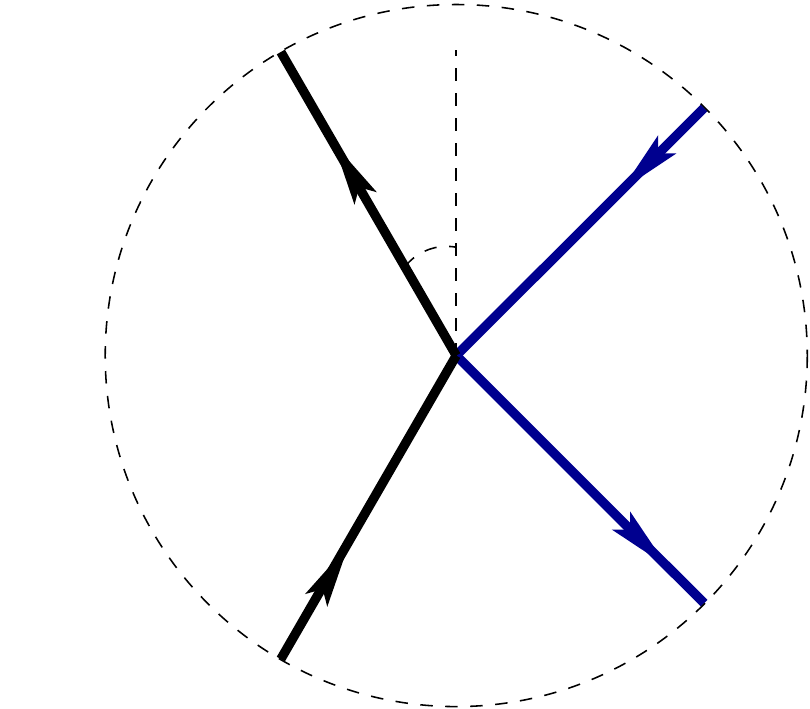} 
\hspace{-10pt}
\resizebox{0.4\textwidth}{!}{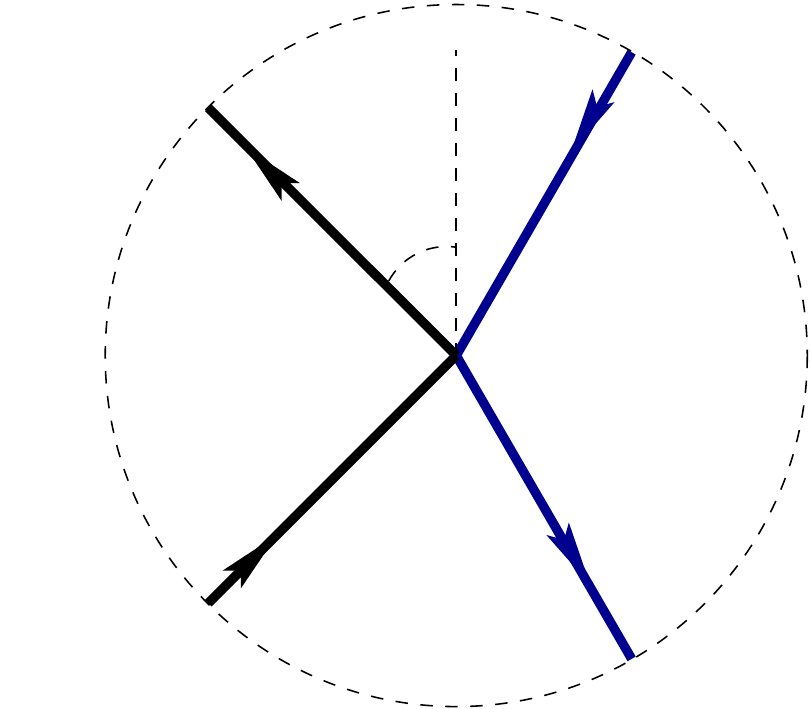}
\begin{minipage}{0.4\textwidth}
\caption{The exact jumps of $\wt \Gamma_{\Ai}$ near $z=1$.
}
\label{PsitildeR}
\end{minipage}\hspace{0.15\textwidth} \begin{minipage}{0.4\textwidth}
\caption{The exact jumps of $\wt \Gamma_{\Ai}$ near $z=-1$.
}
\label{PsitildeL}
\end{minipage}
\end{figure}

\bea
\wt\Gamma_{\Ai}(z) :=\le[{\rm e}^{\frac 1 8 \lambda_R^4\sigma_3}   \Gamma_{\Ai}(\zeta_R;\rho){\rm e}^{ - \frac 1 8 \lambda_R^4\sigma_3} \ri] \\
\lambda_R:= \lambda-\Lambda^3 = \Lambda^{3}(z-1)\ ,\ \ 
\zeta_{_R}:=\Lambda {\lambda_R}{\sqrt[3]{3}}
\eea

It solves the  jumps within the disk centered at $\Lambda^3$ as indicated in Fig. \ref{PsitildeR} (see Remark \ref{rem36}). Its behavior on the boundary $\Lambda^3|z-1| = |\lambda_R|=1$ is 

\be
\wt \Gamma_{\Ai}(z) = \1 +\frac 1{\Lambda \lambda_{_R}\sqrt[3]{3}} \le[
\begin{array}{cc}
p(\rho) & 
{i}
 {\rm e}^{\frac 14 \lambda_{_R}^4} q(\rho)\\
{-i}{\rm e}^{ - \frac 14 \lambda_{_R}^4} q(\rho) & -p(\rho)
\end{array}
\ri] + \mathcal O(\Lambda^{-2})\nonumber 
\ee
With that in mind we shall define the parametrix as follows

\bea
\mathcal P_{R}(z):={\rm diag}(0,0,1) &\& + \le[\le(\1 - \frac 1{\Lambda \lambda_{_R}\sqrt[3]{3}} \le[
\begin{array}{cc}
0 &
{i}
 ({\rm e}^{\frac 14 \lambda_{_R}^4} - 1) q(\rho)\\
{-i}
({\rm e}^{ - \frac 14 \lambda_{_R}^4}-1 )q(\rho)  & 0
\end{array}
\ri] \ri)\wt \Gamma_{\Ai}(z)\ri]^{(1,2)}
\label{549}
\\
\mathcal P_R(z)\bigg|_{|z-1|=\Lambda^{-3}}  &\& = \1 +  \frac 1{ \Lambda^4 (z-1)\sqrt[3]{3}}\le[
\begin{array}{ccc}
p(\rho) &  
{i}
q(\rho)&0\\
{-i}
 q(\rho)&-p(\rho)&   0\\
 0& 0 & 0\\
\end{array}
\ri] + \mathcal O^{(1,2)} \le (\frac{\sqrt{\rho}}{\Lambda^2} {\rm e}^{-\frac 32 \rho^{\frac 23}}\ri ) \label{542}
\eea
The notation $ \mathcal O^{(1,2)} (\Lambda^{-2} {\rm e}^{-\frac 32 \rho^{\frac 23}})$ denotes a matrix of the indicated order only in the entries $i,j\in \{1,2\}$ and zero elsewhere (the bound as a function of $\rho$ follows from the estimate (\ref{smallPII})).
We thus have 

\begin{prop}
\label{proppr}
The matrix $\mathcal P_R$ fulfills the requirements:
\begin{enumerate}
\item is bounded in the disk $|z-1|<\Lambda^{-3}$ ($|\lambda_R|<1$);
\item within the same disk it solves exactly the jumps conditions with $\wh M_R$ replaced by $\wh M_R^{(0)}$ as defined in  (\ref{538});
\item on the boundary $|\lambda_R|=1$ has behavior (\ref{542});
\item the determinant $\det \mathcal P_R$ tends to one uniformly within the disk, and hence $\mathcal P_R^{-1}$ is bounded in the same disk.
\end{enumerate}
\end{prop}
{\bf Proof.}
Only the first and last point require comment while the remaining ones follow from direct manipulations of the definitions.
The boundedness follows from the boundedness of $\wt\Gamma_{\mathrm{Ai}}$ and the boundedness of the prefactor in (\ref{549}); indeed the apparent pole for $z=1 \ \Leftrightarrow\ \lambda_R=0$ is not present  since $\frac 1 {\lambda_R} ({\rm e}^{\pm \frac 1 4 \lambda_R^4}-1)$ are analytic at $\lambda_R=0$ (i.e. $z=1$). 
As for the determinant 
\be
\det\mathcal P_L(z) = 1 +  \frac 1{\Lambda^2}\frac {2\sinh(\lambda_R^4/8)q^2(\rho)}{3 \lambda_R^2} 
\ee
The function $w^{-2} \sinh(w^4/8)$ is bounded in the disk $|w|<1$ (which is our disk in the $z$--plane since $\lambda_R = \Lambda^4(z-1)$ and $|z-1|\leq \Lambda^{-3}$). On the other hand the function $q(\rho)$ is uniformly bounded in the range $K_1<\rho<\frac {K_2}{\sqrt[3]{3}} \Lambda^8$ because $q(\rho)$ is continuous on $\R$ (it is the Hastings-McLeod solution of PII) and decays to zero at $\rho\to+\infty$ as the Airy function $\Ai(\rho)$. Thus the assertion (since there is a division by $\Lambda^2$).
\QED

\subsubsection{Parametrix near \texorpdfstring{$z=-1$}{z-1}}

Similarly to the case $z=1$,consider the  matrix 
\bea
\wt\Gamma_{\mathrm{Ai}}(z) :=\le[{\rm e}^{\frac 1 8 \lambda_L^4\sigma_3}  \sigma_2^{-1} \Gamma_{\mathrm{Ai}}(\zeta_{L};\sigma)\sigma_2 \,{\rm e}^{ - \frac 1 8 \lambda_L^4\sigma_3} \ri] \cr
\lambda_L:= \lambda+\Lambda^3 = \Lambda^3(z+1) \ ,\ \  \zeta_{_L}:= {\lambda_L}{\sqrt[3]{3}\Lambda}\cr
\mathfrak \sigma_2:= \le[\matrix{0&-i\cr i &0}\ri]
\eea

It solves the RHP with the jumps indicated in Fig. \ref{PsitildeL}. Its behavior on the boundary $|z+1| =\Lambda^{-3}$ ( $|\lambda_L|=1$) is 
\be
&\wt \Gamma_{\mathrm{Ai}}(\lambda) = \1 +\frac 1{\Lambda \lambda_{_L}\sqrt[3]{3}} \le[
\begin{array}{cc}
-p(\sigma) &  
-
{i}
{\rm e}^{\frac 14 \lambda_{_L}^4} q(\sigma)\\
{i}
{\rm e}^{ - \frac 14 \lambda_{_L}^4}q(\sigma) & p(\sigma)
\end{array}
\ri] +\mathcal O(\Lambda^{-2})\ .\nonumber 
\ee
With that in mind we shall define the parametrix as follows
\bea
\mathcal P_{L}(z):={\rm diag}(0,1,0)+ \le[\le(\1 - \frac 1{\Lambda \lambda_{_L}\sqrt[3]{3}} \le[
\begin{array}{cc}
0 &
{i}
 ({\rm e}^{\frac 14 \lambda_{_L}^4} - 1)q(\sigma)\\
{-i}
({\rm e}^{ - \frac 14 \lambda_{_L}^4} -1)q(\sigma)  & 0
\end{array}
\ri] \ri)\wt \Gamma_{\mathrm{Ai}}(z)\ri]^{(1,3)}\nonumber
\eea\bea
\mathcal P_L(z)\bigg|_{|z+1|=\Lambda^{-3}}  = \1 +  \frac 1{\Lambda^4 (z+1) \sqrt[3]{3}}\le[
\begin{array}{ccc}
-p(\sigma) &0& -
{i}
q(\sigma) \\
0&0&0\\
{i}
q(\sigma)&0&  p(\sigma)\\
\end{array}
\ri] + \mathcal O^{(1,3)} \le(\frac{\sqrt{\s}}{\Lambda^2} {\rm e}^{-\frac 32 \sigma^\frac 23}\ri)
\label{550}\eea
The next proposition is proved as in the previous section

\begin{prop}
\label{proppl}
The matrix $\mathcal P_L$ fulfills the requirements:
\begin{enumerate}
\item is bounded in the disk $|z+1|<\Lambda^{-3}$ ($|\lambda_L|<1$);
\item within the same disk it solves exactly the jumps conditions with $\wh M_L$ replaced by $\wh M_L^{(0)}$ (\ref{539});
\item on the boundary $|z+1| = \Lambda^{-3}$ ($|\lambda_L|=1$) has behavior (\ref{550});
\item the determinant $\det \mathcal P_L$ tends to one uniformly within the disk, and hence $\mathcal P_L^{-1}$ is bounded in the same disk.
\end{enumerate}
\end{prop}

\subsection{Approximation and error term for the matrix \texorpdfstring{ $Y$}{zeta}} 

We will define  
\bea
\Phi(z):= \le\{
\begin{array}{cc}
\ds \Phi_0(z ):= \1 + \frac {F_1(\rho)^{(1,2)} }{\sqrt[3]{3} \Lambda^4 (z-1)} + 
\frac {(\sigma_2^{-1} F_1(\sigma) \sigma_2)^{(1,3)} }{\sqrt[3]{3} \Lambda^4 (z+1)} & |z-1|>\Lambda^{-3}<|z+1|\\[12pt]
\mathcal P_{R,L }(\lambda ) & |z\mp 1|\leq \Lambda^{-3}
\end{array}
\ri.
\label{449}
\eea
where $F_1(s)$ was introduced in eq.  (\ref{F1}).

\begin{wrapfigure}{r}{0.3\textwidth}
\resizebox{0.3\textwidth}{!}{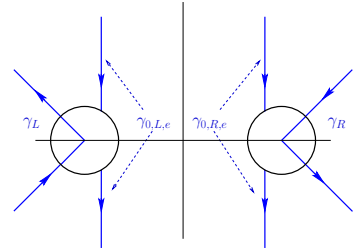}
\caption{The jumps of the error term $\mathcal E$.}
\label{ERHP}
\end{wrapfigure}

\begin{lemma}
\label{Ejumps}
For $K_1<\rho, \sigma< K_2 3^{-\frac 13} \Lambda^8$ , $K_1>-\infty, \ K_2<2$ we have that the error matrix  $\mathcal E (z):= Y(z) \Phi^{-1}(z)$  solves a  Riemann--Hilbert problem with jumps on the contours indicated in Fig. \ref{ERHP} and of the following orders  
\bea
\mathcal E_+(z) =  &\&\mathcal E_-(z) J_{R,L}(z)  \ ,\qquad |z\pm 1|=\Lambda^{-3}\cr
 J_{R,L}(z) :=&\&\Phi_0(z) \mathcal P_{R,L}^{-1}(z)\\
 J_{L}(z) = \1 +&\& \mathcal O^{(1,3)}\!\!\le(\frac {\sqrt{\s} {\rm e}^{-\frac 23 \sigma^\frac 32}}{\Lambda^2} \ri)\!\! +
 \!\!
  \mathcal O^{(1,2)}\!\! \le(\!\frac{\sqrt{\rho}{\rm e}^{-\frac 23 \rho^\frac 32}}{\Lambda^4}\!\ri)
\!\!  +\!\! \mathcal O\!\!\!\le (\!\frac{\sqrt{\rho\s} {\rm e}^{-\frac 32 (\rho^\frac 23 + \sigma^\frac 23)}}{\Lambda^{5} }\! \ri)\!\!;  |z+1|=\Lambda^{-3}
  \label{555}\\ 
 J_{R}(z)= \1 + &\&  \mathcal O^{(1,3)}\!\!\le(\! \frac {\sqrt{\s} {\rm e}^{-\frac 23 \sigma^\frac 32}}{\Lambda^4} \!\ri)\!\! +
 \!\!
  \mathcal O^{(1,2)}\!\! \le(\!\frac{\sqrt{\rho}{\rm e}^{-\frac 23 \rho^\frac 32}}{\Lambda^2}\!\ri)
\!\!  +\!\! \mathcal O\!\!\!\le (\!\frac{\sqrt{\rho\s} {\rm e}^{-\frac 32 (\rho^\frac 23 + \sigma^\frac 23)}}{\Lambda^{5} }\! \ri)\!\! 
; |z-1| = \Lambda^{-3}
 \label{554}
\eea
whereas on the remainder of the contours we have
\be
\mathcal E_+(z) = \mathcal E_-(z) J_O(z)\ ,\ \ J_O(z)= \1 + \mathcal O(\Lambda^{-\infty}).
\ee
and the big-$\mathcal O$ symbols are in any $L^p$ norms ($p\leq \infty$). The subscript on $J_O$ stands for ''Other''.
\end{lemma}
{\bf Proof.}
On the boundaries of the disks the estimates (\ref{554}, \ref{555}) (including the shape) follows from (\ref{smallPII}) and (\ref{542}, \ref{550}), the definition of $\Phi_0$ (\ref{449}) and matrix multiplication.

The other jumps are on $\gamma_{0,R,e}, \gamma_{0,L,e},\gamma_L, \gamma_R$;  on $\gamma_{0,R,e}, \gamma_{0,L,e}$ the jumps are $\Phi_0 \wh M_{0,R} \Phi_0^{-1}$, $\Phi_0 \wh M_{0,L} \Phi_0^{-1}$ and Lemma \ref{Lemma3} together with the fact that both $\Phi_0, \Phi_0^{-1}$ are uniformly bounded implies that the jump is within $\mathcal O({\rm e}^{-\Lambda^3 - \frac {K_1}2 \Lambda})$ of the identity in any $L^p$, $p\leq \infty$.
On $\gamma_{R}$ and outside the right disk the jump is  $\Phi_0 \wh M_{R} \Phi_0^{-1}$ and now Lemmas \ref{Lemma1} and \ref{Lemma2} imply similarly that the jumps are within $\mathcal O(\Lambda^{-\infty})$  of the identity.  On $\gamma_R$ {\em inside} the disks, since the jump of $\mathcal P_R$ is $\wh M^{(0)}_R$ (Prop. \ref{proppr}, point 2), the jump of $\mathcal E$  is $\mathcal P_{R-} \wh M_R (\wh M^{(0)}_R)^{-1} \mathcal P_{R-}^{-1}$. Comparing with (\ref{538}) and using the boundedness of $\mathcal P_R, \mathcal P_R^{-1}$ (Prop. \ref{proppr}, point 4) we see that the jump is also exponentially close to the identity. Similarly for the jump on $\gamma_L$ outside and within the left disk.
\QED

The small--norm theorem (which we sketch in App. \ref{smallnorm}) says  that --uniformly on closed sets not containing the contours of the  jumps which we collectively denote with $\gamma$--
\be
\le\|\mathcal E(z) -\1 \ri\| = \frac 1{{\rm dist}(z,\gamma)} \mathfrak C(J)
\label{errorestim}
\ee
where the constant $\mathfrak C(J)$  is bounded by some multiple of the $L^1$ norm and the {\bf square} of the $L^2$ norm of $J(z)-\1$ (here $J(z)$ is the jump on all possible contours for $\mathcal E$).  Both of these two quantities on a circle (or any compact contour for that matter) are bounded by a multiple of the length of the curve. 
Overall the main contribution to $\mathfrak C(J)$ is coming from the two circles and since these shrink as $\Lambda^{-3}$  (so does their length) we have 
\be
\mathfrak C(J) \leq Const\, \Lambda^{-3} \|J(z)-\1\|_{L^\infty}
\ee
with the $L^\infty$ norm here above restricted to the two shrinking circles.
The error matrix $\mathcal E(z)$ is then found as the solution of the singular-integral equation 
\be
\mathcal E(z) = \1 + \int \frac {\mathcal E_-(w) (J(w)-\1) \d w}{2i\pi(w-z)}
\ee
and can be obtained by iterations 
\be
\mathcal E^{(0)} \equiv 1\ ,\ \ \mathcal E^{(k+1)}(z):=  \1 + \int \frac {\mathcal E_-^{(-1)} (w) (J(w)-\1) \d w}{2i\pi(w-z)}
\ee
where the integral extends to all the contours of the jumps; however the only contributions that are significant come from the boundaries of the two disks (the other contributions give exponentially small terms).
This way one promptly verifies that the matrix $\mathcal E(z)$  has orders in the following shape
which follows from the particular shape of the jumps $J_L, J_R$ in (\ref{555}, \ref{554})
\be
\mathcal E(z) &\& = \1 + \frac {1}{ {\rm dist}(z,\gamma)}
\mathcal O_{\rho,\sigma}\cr
&\& \mathcal O_{\rho,\sigma}:= \mathcal O^{(1,2)}
\le(\frac {\sqrt{\rho}}{\Lambda^5} {\rm e}^{-\frac 32 \rho^\frac 23}\ri)
 +
  \mathcal O^{(1,3)}\le (\frac{\sqrt\s}{\Lambda^5} {\rm e}^{-\frac 32 \sigma^\frac 23}\ri)
+
\mathcal O\le(
\frac{\sqrt{\rho\s}}{\Lambda^5} {\rm e}^{-\frac 32 (\sigma^\frac 23   + \rho^\frac 23)} \ri) 
\label{Eestim}
\ee

\subsection{Conclusion of the proof of  Theorem \ref{main1}}
By applying Thm. \ref{FredholmtauII} we have 
 for $a_{-}=a_{-}(\rho),\ a_{+}=a_{+}(\sigma)$ with $a_\pm$ given in (\ref{apm})
\bea
\pa_{a_{-}}\ln \det\le(\Id - K_{_P}\bigg|_{[a_{-},a_{+}]}\ri)  = -\Gamma_{1;2,2}\  ,\ \ \  
\pa_{a_{+}}\ln \det\le (\Id - K_{_P}\bigg|_{[a_{-},a_{+}]}\ri)  =- \Gamma_{1;3,3}
\eea
where $\Gamma_1:= \lim_{\lambda\to \infty} \lambda (\Gamma(\lambda) -\1)$;
to compute the limit we take $\lambda$ within the sector $\frac \pi 4+\epsilon<\arg(\lambda)< \frac \pi 2-\epsilon$ (form some $\epsilon>0$) so that the distance between $z= \lambda \Lambda^{-3}$ and the jumps of $\mathcal E$ grows like $\mathcal O(z)$. then our approximation for $\Gamma(\lambda)$ from (\ref{Eestim})
\be
\Gamma(\lambda) &\& = {\rm e}^{D_0} \mathcal E\le(\frac {\lambda}{\Lambda^3}\ri) \Phi_0\le(\frac {\lambda}{\Lambda^3} \ri) {\rm e}^{-D_0} =\nonumber \\
&\& = 
{\rm e}^{D_0} \le( \1 + \frac{\mathcal O_{\rho,\sigma}}{ \frac \lambda{\Lambda^3}}  + \frac {F_1(\rho)^{(1,2)} }{\sqrt[3]{3} \Lambda (\lambda-\Lambda^3)} + 
\frac {(\sigma_2^{-1} F_1(\sigma) \sigma_2)^{(1,3)} }{\sqrt[3]{3} \Lambda (\lambda+\Lambda^3)} +\mathcal O(\lambda^{-2})   \ri){\rm e}^{-D_0} 
\ee
Thus, computing the limit that defines the matrix $\Gamma_1 =  \lim_{\lambda\to \infty} \lambda (\Gamma(\lambda) -\1)$ we have 
\be
\Gamma_1 = {\rm e}^{D_0} \le(
 \mathcal O^{(1,2)}\le(\frac { \sqrt\rho {\rm e}^{-\frac 32 \rho^\frac 23}}{\Lambda^2} \ri) 
 +
 O^{(1,3)}\le(\frac { \sqrt\s {\rm e}^{-\frac 32 \sigma^\frac 23}}{\Lambda^2} \ri) 
 +
  \mathcal O\le(\frac { \sqrt{\rho\s}{\rm e}^{-\frac 32 (\rho^\frac 23+ \sigma^\frac 23) }}{\Lambda^2} \ri) 
+\ri.
\nonumber
\\
\le.  + \frac {F_1(\rho)^{(1,2)} }{\sqrt[3]{3} \Lambda} + 
\frac {(\sigma_2^{-1} F_1(\sigma) \sigma_2)^{(1,3)} }{\sqrt[3]{3} \Lambda}
\ri) {\rm e}^{-D_0}
\label{557}
\ee
From (\ref{557})  we have
\bea
\Gamma_{1;2,2}&\& =
  -\frac 1{ \sqrt[3]{3}\Lambda} p(\rho) + \mathcal O\le(\frac{\sqrt\rho {\rm e}^{-\frac 32 \rho^\frac 23}}{\Lambda^2}\ri) 
  +\mathcal O\le(\frac { \sqrt{\rho\s}{\rm e}^{-\frac 32 (\rho^\frac 23+ \sigma^\frac 23) }}{\Lambda^2} \ri) \\  
\Gamma_{1;3,3}  &\& =
  \frac 1{ \sqrt[3]{3}\Lambda}p(\sigma)+
   \mathcal O \le(\frac{\sqrt{\s}{\rm e}^{-\frac 32 \sigma^\frac 23}}{\Lambda^2}\ri) 
  +\mathcal O\le(\frac {\sqrt{\rho\s} {\rm e}^{-\frac 32 (\rho^\frac 23+ \sigma^\frac 23) }}{\Lambda^2} \ri)
\eea

Recall that  $p(\sigma)$ is the logarithmic derivative of the gap probability for the Airy process;
noticing that $\pa_{a_-}  = \frac 1{\sqrt[3]{3}\Lambda} \pa _\rho,\ \pa_{a_+}  =- \frac 1{\sqrt[3]{3}\Lambda} \pa _\sigma$, the above equations say (in the notation of differential forms)
\bea
\d_{\rho,\sigma}&\&\ln \det\le(\Id - K_{_P}\bigg|_{[a_{-}(\rho), a_{+}(\sigma)]}\ri)= \d_\sigma  \ln \det\le(\Id - K_{\mathrm{Ai}}\bigg|_{[\sigma,\infty)} \ri)  + 
\d_\rho  \ln \det\le(\Id - K_{\mathrm{Ai}}\bigg|_{[\rho,\infty)} \ri) +\cr
&\& +   \mathcal O\le(\frac{\sqrt{\rho}}{\Lambda}{\rm e}^{-\frac 32 \rho^\frac 23}\ri) \d \rho + 
\mathcal O \le(\frac{\sqrt{\s}}{\Lambda}{\rm e}^{-\frac 32 \sigma^\frac 23}\ri) \d\sigma 
  + \mathcal O\le(\frac { \sqrt{\rho\s}{\rm e}^{-\frac 32 (\rho^\frac 23+ \sigma^\frac 23) }}{\Lambda} \ri)(\d \rho + \d \sigma)
 \label{452}
\eea
where we recall that the estimates hold uniformly for $K_1<\rho, \sigma <K_2 3^{-1/3} \Lambda^8$  and $0<K_2<2,\ K_1>-\infty$.
To obtain information in integral form we integrate (\ref{452}) from  some point $(\rho_0, \sigma_0)$ to the point $(\rho,\sigma)=\Lambda^8/2(1,1)$; this point is still within the range of validity of all of our estimates in Lemmas \ref{Lemma1}, \ref{Lemma2}, \ref{Lemma3} (in particular Lemma \ref{Lemma2}) and corresponds to an interval $[a_{-},a_{+}]$ of fixed size (not growing as $\tau=3 \Lambda^6 \to\infty$) containing the origin in its interior. The correction terms in the differential form (\ref{452}) yield a form that  is integrable on  any rectangular domain of the form  
\be
K_1 \leq \rho,\sigma \leq \frac {K_2}{\sqrt[3]{3}} \Lambda^8,\ \ -\infty<K_1, \ K_2<2.
\ee
and the result is a function of order $\mathcal O(\Lambda^{-1})$ thanks to (\ref{452}). 
Thus we conclude 
\be
\ln\! \det\!\!\le(\Id - K_{_P}\bigg|_{[a_{-}(\rho), a_{+}(\sigma)]}\ri)\!\!=\!\!\ln\! \det\!\!\le(\Id - K_{\mathrm{Ai}}\bigg|_{[\sigma,\infty)} \ri)\!\! + 
 \ln\! \det\!\!\le(\Id - K_{\mathrm{Ai}}\bigg|_{[\rho,\infty)} \ri)\!\! +\mathcal O(\Lambda^{-1}  ) + C\nonumber
\ee
 It is shown in App. \ref{appB} that the Fredholm determinant of the Pearcey kernel in this regime (i.e. when the interval $[a_-,a_+]$ is fixed) tends exponentially fast to unity as $3\Lambda^6=\tau \to \infty$.  
On the other hand also the Fredholm determinants of the Airy-kernels on the right-hand side tend to unity, and hence the constant of integration $C$ must be zero.
The proof of Thm. \ref{main1} is now complete. \QED

The proof of Thm. \ref{main2} is completely analogous, with the complication that  the local parametrices would have to be built out of bigger-size version of the Hastings-McLeod auxiliary matrix constructed using Def. \ref{multiAi} and the analogue of the simple estimates in Section. \ref{auxiliarymatrix} for this larger-size version would have to be derived.

\appendix
\renewcommand{\theequation}{\Alph{section}.\arabic{equation}}

\section{Proof of Theorem \ref{thdetMalgrange}}
\label{AppA}
We briefly recall the situation we already described in the introduction. Following \cite{ItsHarnad} (and references therein), we start with an integrable kernel $K(\lambda,\mu )$ acting on $L^2(\cal C)$ where $\cal C$ is a analytic contour in $\C$, possibly extending to infinity. Being integrable means that we can write the kernel as
$$K(\lambda,\mu ):=\frac{\f^{\mathrm T}(\lambda)\g(\mu )}{\lambda-\mu }$$ 
with $\f^{\mathrm T}(\lambda)\g(\lambda)=0$ (i.e. the kernel is non-singular on the diagonal). Moreover we assume that vectors $\f$ and $\g$ are analytic in a neighborhood of $\cal C$. Then the resolvent  $R=(\Id-K)^{-1}K$ is again an integrable operator related to the vectors $\F,\G$ given by
 \bea
        		\vec F(\lambda)= \Gamma(\lambda)\vec f(\lambda)\ ,\qquad 
		\vec G(\lambda) = (\Gamma^{-1})^{\mathrm T}(\lambda)\vec g(\lambda)
\eea
where $\Gamma$ is the solution of the RH problem (\ref{RH1})-(\ref{RH2})-(\ref{RH3}). From now on we denote with  a prime  the derivative w.r.t. the spectral parameter. 

Using formula (\ref{resolvent}) we have
$$\partial\log(\det(\Id-K))=-\Tr(\partial K+R\partial K).$$
The first piece is immediately computable since $K(\lambda,\lambda)=\vec f'^\T(\lambda)\vec g(\lambda)$ so that
\be
	-\Tr(\partial K)=-\int_{\cal C} \Big(\partial\vec f'^\T(\lambda)g(\lambda)+\vec f'^\T(\lambda)\partial\vec g(\lambda)\Big)d\lambda=H_1(M).
\label{eq0}\ee
For the other part we get, using the explicit expression of the resolvent,
\bea
	-\Tr(R\partial K)=&\&
	\int\int_{\cal C}\frac{-1}{(\lambda-\mu )^2}\Tr\Big((\g^\T\Gamma^{-1})(\mu )(\Gamma\f)(\lambda)\big(\partial\g^\T(\lambda)\f(\mu )+\g^\T(\lambda)\partial\f(\mu )\big)\Big)d\lambda d\mu \nonumber\\
	=
	\int\int_{\cal C} &\&\frac{-1}{(\lambda-\mu )^2}\Tr\Big(\big(\f\g^\T\Gamma^{-1}\big)(\mu )\big(\Gamma\f(\partial\g^\T)\big)(\lambda)+\big((\partial\f)\g^\T\Gamma^{-1}\big)(\mu )\big(\Gamma\f\g^\T\big)(\lambda)\Big)d\lambda d\mu \nonumber\\
	=
	\frac{1}{2\pi i}\int\int_{\cal C} &\&\Tr\Big(\big(\Gamma_-^{-1}-\Gamma_+^{-1}\big)(\mu )\big(\Gamma\f(\partial\g^\T)\big)(\lambda)+\Gamma^{-1}(\mu )\big(\Gamma_+-\Gamma_-\big)(\lambda)\big((\partial\f)\g^\T\big)(\mu )\Big)\frac{ d\lambda d\mu }{(\lambda-\mu )^2}\nonumber
\eea
where we just used the invariance of the trace for cyclic permutations and, in the last passage, we used the two identities
\bea
	(\Gamma_+-\Gamma_-)(\lambda)&=&{-}
	2\pi i(\Gamma\f\g^\T)(\lambda)\label{identity1}\\
	(\Gamma_+^{-1}-\Gamma_-^{-1})(\lambda)&=&
	2\pi i(\f\g^\T\Gamma^{-1})(\lambda)\label{identity2}
\eea
coming directly from the related RH problem.

Now we work with the assumption that $\f,\g$ are analytic in a neighborhood of the contour and we split the integral above in two part. The two parts will be singular when $\lambda=\mu $ so that we have to pay attention to the order of integration. More precisely we write it as
\bea
	-\Tr(R\partial K)&=&
	\int\int_{\cal C} \frac{1}{(\lambda-\mu )^2}\Tr\Big(\big(\Gamma_-^{-1}-\Gamma_+^{-1}\big)(\mu )\big(\Gamma\f(\partial\g^\T)\big)(\lambda)\frac{d\lambda}{2\pi i}\Bigg|_{\mu =\mu _-}d\mu \nonumber\\
	&{+}&\int\int_{\cal C} \frac{1}{(\lambda-\mu )^2}\Tr\Big(\Gamma^{-1}(\mu )\big(\Gamma_+-\Gamma_-\big)(\lambda)\big((\partial\f)\g^\T\big)(\mu )\Big)\frac{d\lambda}{2\pi i}\Bigg|_{\mu =\mu _-}d\mu \label{eq1}
\eea
We would like to apply Cauchy Residue's theorem but, in the first term, we have to interchange the order of integration. Given an analytical function $F(\lambda,\mu )$ in a neighborhood of the contour we observe that we have
\bea
	\int\int_{\cal C} \frac{1}{(\lambda-\mu )^2}F(\lambda,\mu )\frac{d\lambda}{2\pi i}\Bigg|_{\mu =\mu _-}d\mu &=&\int\int_{\cal C} \frac{1}{(\lambda-\mu )^2}F(\lambda,\mu )\frac{d\mu }{2\pi i}\Bigg|_{\lambda=\lambda_+}d\lambda\nonumber\\
	&=&\int\int_{\cal C} \frac{1}{(\lambda-\mu )^2}F(\mu ,\lambda)\frac{d\lambda}{2\pi i}\Bigg|_{\mu =\mu _+}d\mu \nonumber
\eea
where in the last equation we just renamed the dummy variables. Hence using Cauchy's theorem and the equation above for the first term of the r.h.s. of (\ref{eq1}) we get 
\bea
	-\Tr(R\partial K)= 
	{-}
	\int_{\cal C} \Tr\Big(\big(\Gamma_+^{-1}\Gamma_+'\f(\partial \g)^\T\big)(\mu )\Big)d\mu 
	{-}
	\int_{\cal C} \Tr\Big(\big(\Gamma_-^{-1}\Gamma_-'(\partial\f)\g^\T\big)(\mu )\Big)d\mu \label{eq2}
\eea
Hence, combining (\ref{eq0}) and (\ref{eq2}), we get
\bea
	\partial\log(\det(\Id-K))=-\Tr(\partial K+R\partial K)=\cr=
	H_1(M)
	{-}
	\int_{\cal C} \Tr\Big(\big(\Gamma_+^{-1}\Gamma_+'\f(\partial \g)^\T+
	\Gamma_-^{-1}\Gamma_-'(\partial\f)\g^\T\big)(\mu )\Big)d\mu 
\label{eq2bis}
\eea

On the other hand we have that
\bea
	\omega_M(\partial)=\int_{\cal C}\Tr\Big(\Gamma_-^{-1}\Gamma_-'\partial M M^{-1}\Big)\frac{d\lambda}{2\pi i}&\&= 
	{-}
	\int_{\cal C}\Tr\Big(\Gamma_-^{-1}\Gamma_-'((\partial \f)\g^\T+\f(\partial\g^\T))(\1
	{+}
	2\pi i\f\g^\T)\Big)d\lambda 
	\nonumber\\
	= {-}
	\int_{\cal C}\Tr\Big(\Gamma_-^{-1}&\&\Gamma_-'((\partial \f)\g^\T+\f(\partial\g^\T))
	{+}
	2\pi i\Gamma_-^{-1}\Gamma_-'\f(\partial\g)^\T\f\g^\T)\Big)d\lambda
	\nonumber\\
	=
	{-}
	\int_{\cal C} Tr\Big(\Gamma_-^{-1}&\&\Gamma_-'((\partial \f)\g^\T+\f(\partial\g)^\T)+(\Gamma_+^{-1}-\Gamma_-^{-1})\Gamma_-'\f(\partial\g)^\T)\Big)d\lambda\nonumber\\
	={-}
	 \int_{\cal C}\Tr\Big(\Gamma_-^{-1}&\&\Gamma_-'(\partial \f)\g^\T+\Gamma_+^{-1}\Gamma_-'\f(\partial\g)^\T)\Big)d\lambda\label{eq3}
\eea
(here again we used (\ref{identity2})). Now we observe that, because of the RH problem, we have
\be
\Gamma_-'=\Gamma_+'M^{-1}
{+}
2\pi i\Gamma_+(\f\g^\T)'
\ee 
so that
\bea
	\int_{\cal C}\Tr(\Gamma_+^{-1}\Gamma_-'\f(\partial\g)^\T)d\lambda
	&=&\int_{\cal C}\Tr\Bigg(\Gamma_+^{-1}\Big(\Gamma_+'(\1
	{+}
	2\pi i\f\g^\T)
	{+}
	2\pi i\Gamma_+(\f\g^\T)'\Big)\f(\partial\g)^\T\Bigg)d\lambda\nonumber\\
	&=& \int_{\cal C} \Tr\Big(\Gamma_+^{-1}\Gamma_+'\f(\partial\g)^\T\Big)d\lambda
	{-}
	2\pi i\int_{\cal C}\Tr(\f\g^\T\f'(\partial\g)^\T)d\lambda\label{eq4}
\eea
where we used several times the invariance of the trace under cyclic permutations and the fact that $\f^\T(\lambda)\g(\lambda)\equiv 0$. Substituting (\ref{eq4}) into (\ref{eq3}) we get
\bea
	\omega_M(\partial)=&\&
	\int_{\cal C}\Tr\Big(\Gamma_-^{-1}\Gamma_-'\partial M M^{-1}\Big)\frac{d\lambda}{2\pi i}=\nonumber\\
	=&\&
	{-}
	\int_{\cal C}\Tr\Big(\Gamma_-^{-1}\Gamma_-'(\partial \f)\g^\T+\Gamma_+^{-1}\Gamma_+'\f(\partial\g)^\T\Big)d\lambda
	+
	2\pi i\int_{\cal C}\Tr(\f\g^\T\f'(\partial\g)^\T)d\lambda=\nonumber\\
	=&\&
	{-}
	\int_{\cal C}\Tr\Big(\Gamma_-^{-1}\Gamma_-'(\partial \f)\g^\T+\Gamma_+^{-1}\Gamma_+'\f(\partial\g)^\T\Big)d\lambda+H_2(M)\label{eq5}
\eea
Comparing (\ref{eq5}) with (\ref{eq2bis}) we obtain (\ref{detMalgrange}). \QED

\begin{remark}
It can also be proven that 
$$
H_2(M)=-\frac{1}2\int_{\cal C}\Tr(M'M^{-1}{\partial M} M^{-1})d\lambda
$$ 
through the following computation
\bea
	\int_{\cal C}\Tr(M'M^{-1}(\partial M)M^{-1})\frac{d\lambda}{2\pi i}&\& =
	\int_{\cal C}\Tr\Big((\f\g^\T)'(\1
	{+}
	2\pi i\f\g^\T)\partial(\f\g^\T)(\1
	{+}
	2\pi i \f\g^\T)\Big)d\lambda\nonumber\\
	= \int_{\cal C}\Tr\Big(\f'\g^\T(\partial\f)\g^\T&\& +\f\g^{\T'}\f(\partial \g)^\T\Big)d\lambda\nonumber\\
	= \int_{\cal C}\Tr\Big(\g^\T\f'\g^\T(\partial \f)&\& -\f\g^\T\f'(\partial \g)^\T\Big)d\lambda=-\int_{\cal C}\Tr\Big(\g^\T\f'(\partial \g)^\T\f+\f\g^\T\f'(\partial \g)^\T\Big)d\lambda\nonumber\\
	= -2\int_{\cal C}\Tr\Big(\f\g^\T\f'(\partial &\&g)^\T \Big)d\lambda\nonumber
\eea
This same quantity is introduced in \cite{BertolaIsoTau} and is nothing but the difference between $\omega_M$ and the so-called modified Malgrange's form (see the cited paper for details).

\end{remark}

\section{Behavior of \texorpdfstring{$K_{_P}$}{KP} for \texorpdfstring{$\tau\to+\infty$}{ti}}
\label{appB}
We consider the behavior of the Pearcey kernel when $\tau\to +\infty$ and $x,y$ belong to some fixed bounded interval $[a,b]$. The analysis is a regular case of steepest descent. 

We perform the rescaling $\lambda = \sqrt \tau z$ $ \mu= \sqrt \tau w$ and obtain
\bea
K_{_P}(x,y;\tau) :=\sqrt{\tau} \int_{\gamma_L\cup\gamma_R}\hspace{-10pt} \d w \int_{i\R} \d z \frac {{\rm e}^{\tau^2\vartheta_x(w) -\tau^2 \vartheta_y(z)}}{z - w} \\
\vartheta_x(z):= \frac {z^4}4 - \frac {1}2 z^2 + \frac{x}{\tau^{\frac 32}} z.
\eea
The saddle points for $\Re \vartheta$ are 
\be
\vartheta_x'(z_c) =0  \ \Rightarrow \  z_c= \le\{ \begin{array}{c}
\ds {x}\,{\tau^{-\frac 32}} + \mathcal O (\tau^{-9/2})\\[10pt]
\ds 1-\frac {x}2 { \tau^{-\frac 32}} + \mathcal O (\tau^{-6/2})\\[10pt]
\ds 1-\frac {x}2 { \tau^{-\frac 32}} + \mathcal O (\tau^{-6/2})
\end{array}
\ri. \ \ \vartheta_x(z_c) = \le\{
\begin{array}{c}
\ds -\frac 12 \le(\frac{x^2}{ \tau^{3}} + \mathcal O (\tau^{-9/2})\ri)
\\[10pt]
\ds -\frac 14 \le(1- \frac{x^2}{ \tau^{3}} + \mathcal O (\tau^{-9/2})\ri)\\[10pt]
\ds -\frac 14 \le(1- \frac{x^2}{ \tau^{3}} + \mathcal O (\tau^{-9/2})\ri)
\end{array}
\ri.
\ee
The real part of $\vartheta_x(w)$ has local maxima at $w_c\simeq \pm 1$ in the vertical direction, which is suitable for the contours $\gamma_{L,R}$; viceversa the real part of $-\vartheta_y(z)$ has local max. at $z\simeq 0$ again in the vertical direction, suitable for $i\R$. 
The usual steepest descent asymptotics then gives.
\be
K_{_P}(x,y;\tau) = \frac 1{\sqrt{\tau}}\,{\rm e}^{- \frac 14\tau^2} C(1+\mathcal O(\tau^{-\frac 32}))
\ee
with $C$ a constant (independent of $x,y$) whose value is immaterial for our considerations.
As a consequence of this simple estimate we immediately deduce that 
\be
\le |\ln \det\le (\Id - K_{_P}(\bullet,\bullet;\tau)\bigg|_{[a,b]}\ri)\ri| \leq \sum_{n=1}^{\infty} \frac 1 n (b-a)^n C^n \tau^{-\frac n2} {\rm e}^{-\frac n4 \tau^{2}}\leq \\
\leq \sum_{n=1}^{\infty}  (b-a)^n C^n \tau^{-\frac n2} {\rm e}^{-\frac n4 \tau^{2}}  =  \frac{(b-a) C \tau^{-\frac 12} {\rm e}^{-\frac 14 \tau^{2}}}{1-(b-a) C \tau^{-\frac 12} {\rm e}^{-\frac 14 \tau^{2}}}
\ee
and hence the Fredholm determinant tends (exponentially) to one.
\section{On the small--norm theorem}
\label{smallnorm}
We briefly review the outline of the proof of a ``small norm theorem'' in a somewhat idealized situation to see how the length of the contours affects the estimates. We consider a simplified situation where the jumps are only on the circles (which are anyway the main source of correction terms).
Consider the RHP for $\mathcal E(z)$ on the  a collection of circles  $\Sigma$. For simplicity let's assume  $\Sigma = S_r$ with $S_r:=\{|z|=r\}$. 
\bea
\mathcal E_+(z) = \mathcal E_-(z) (\1 + G(z))\ ,\ \ |z|=r\\
\mathcal E(z) = \1 + \mathcal O(z^{-1})
\eea
$G(z)$ is assumed to be in any $L^p$ of $\Sigma$ and in fact we assume it also  smooth  (this is a very relaxed assumption but it fits our setup and most setups). The solution can be presented as 
\be
\mathcal E(z) = \1 + \oint \frac {\mathcal E_-(w) G(w)\d w}{(w-z)2i\pi}
\ee
Taking the boundary value from the right  we have the following singular integral equation for $\rho(z):= \mathcal E_-(z)-\1\in L^2(S_r)$
\be
\le(Id - \mathcal C_-(\bullet\cdot  G)\ri) \rho = \mathcal C_{-}(G)
\ee
where $\mathcal C_-()$ is the Cauchy boundary value operator on $\Sigma$. On the circle this is the projectors on the analytic part inside the disk, and $\bullet \cdot G$ stands here for the (right) multiplication operator by the matrix $G(z)$. 
It is easy that the operator norm in $L^2(S_r)$ of $\mathcal C_-$  for the circle is one (and independent of the radius!). Therefore the operator norm of $\mathcal L:= \mathcal C_-(\bullet\cdot  G)$ is bounded by 
\be
\||  \mathcal L\|| \leq   \|G\|_{\infty}
\ee
Thus $\mathcal L$ is invertible as long as $ \|G\|_{\infty}$ is less than one and 
\be
\|| (Id - \mathcal L)^{-1}\|| \leq \frac 1{1 -  \|G\|_\infty}
\ee
Thus the solution to the singular integral equation has a bound
\be
\|\rho\|_2 \leq \|| (Id - \mathcal L)^{-1}\|| \,\|\mathcal C_-(G) \|_2 \leq\frac  {\|G\|_{2} }{1 - \|G\|_{\infty}}
\ee
Now  (here $|M|$ is any norm on the space of matrices for example the Hilbert-Schmidt norm $|M|^2 = \Tr (M \ov M^T)$ while $|\d w|$ is the arc-length)
\be
|\mathcal E(z)-\1| \leq  \oint \frac {| (\1 + \rho(w) ) G(w)| |\d w|}{|w-z|2\pi} \leq \frac 1{2\pi {\rm dist}(z,S_r)}  \oint\le( |G(w)| + | \rho(w) G(w)| \ri)|\d w| = \\
=\frac {\|G_1\|_1 + \|\rho\|_2 \|G\|_2}{2\pi {\rm dist}(z,S_r)} \leq \frac 1 {2\pi {\rm dist}(z,S_r)}\le(
\|G\|_1 + \frac{ \|G\|_2^2}{ 1 - \|G\|_\infty}
\ri)  \leq\\
\leq  \frac {\ell(S_r)}  {2\pi{\rm dist}(z,S_r)}\le( \|G\|_\infty + \frac {\|G\|_\infty^2}{1 - \|G\|_\infty} \ri)
\ee
What is important to us here (this is only a rough estimate) is that the estimate is proportional to the length of the circle $\ell(S_r)$. In our problem these are proportional to $\Lambda^{-3}$ and hence they improve the estimate of the error for $z$ in the neighborhood of infinity (which is what we need).
\bibliographystyle{plain}
\bibliography{/Users/bertola/Documents/Papers/BibDeskLibrary.bib}
 \end{document}